\documentclass[prd,preprint,nofootinbib]{revtex4-1}

\usepackage{graphicx}
\usepackage{dcolumn}
\usepackage{bm}

\pdfoutput=1

\usepackage{amssymb}
\usepackage{amsmath}
\usepackage{color}
\usepackage{caption}
\usepackage{subcaption}
\usepackage{hyperref}

\graphicspath{{figures//}}

\begin{document}

\title{Thermal suppression of moving heavy quark pairs in a strongly coupled plasma}
\author{S.~I.~Finazzo}
\email{stefano@if.usp.br}
\author{J.~Noronha}
\email{noronha@if.usp.br}
\affiliation{Instituto de F\'{i}sica, Universidade de S\~{a}o Paulo, S\~{a}o Paulo, SP, Brazil}

\date{\today}

\begin{abstract}
The imaginary part of the heavy quark-antiquark potential experienced by moving heavy quarkonia in strongly coupled plasmas dual to theories of gravity is computed by considering thermal worldsheet fluctuations of the holographic Nambu-Goto string. General results for a wide class of gravity duals are presented and an explicit formula for $\mathrm{Im} \, V_{Q\bar{Q}}$ is found in the case where the axis of the moving $\bar{Q}Q$ pair has an arbitrary orientation with respect to its velocity in the plasma. These results are applied to the study of heavy quarkonia propagating through a strongly coupled $\mathcal{N} = 4$ SYM plasma. Our results indicate that the onset of $\mathrm{Im} \, V_{Q\bar{Q}}$ decreases with increasing rapidity (though our analysis is limited to slowly moving quarkonia) and that, in general, a $Q\bar{Q}$ pair is more strongly bound if its axis is aligned with its direction of motion through the strongly coupled plasma.
\end{abstract}

\maketitle

\section{Introduction}
\label{sec:intro}

In heavy ion collisions, useful probes to studying the formation and evolution of the Quark-Gluon Plasma (QGP) \cite{Shuryak:1980tp} are the heavy quarkonia ($J/\psi$ and $\Upsilon$ mainly) formed in hard processes before the thermalization of the plasma. The seminal work of Matsui and Satz \cite{Karsch:1987pv} argued that, in a thermal bath, the binding interaction of the heavy quark-antiquark ($Q\bar{Q}$)pair is screened by the medium, resulting in the melting of the heavy quarkonia. However, the $Q\bar{Q}$ pair is not necessarily produced at rest in the QGP and the effects of its motion through the plasma must taken into account when considering the effects of the medium in the $Q\bar{Q}$ interaction.

In a non-Abelian $SU(N_c)$ gauge theory in 4 dimensions, a gauge-invariant observable of interest to study the interaction of the medium in the heavy $Q\bar{Q}$ pair interaction is obtained from the Wilson loop operator \cite{Wilson:1974sk,wilsonloop}. The Wilson loop operator $W(C)$ is defined by
\begin{equation}
\label{eq:wilsonloop}
W(C) = \frac{1}{N_c}\textrm{tr} \, P \, \exp{ \left[ i g\oint_C \hat{A}_{\mu} dx^{\mu} \right]},
\end{equation}
where $C$ is a closed loop in spacetime, $g$ is the non-Abelian coupling constant of the gauge field $\hat{A_{\mu}}$ and $P$ is the path-ordering operator. The trace in \eqref{eq:wilsonloop} is over the fundamental representation of the gauge group. In the special case that $C$ is a rectangular loop with two sides along the time $t$ direction and the other two sides along, say, the $x_3$ direction, with the dimensions of the rectangle being $\mathcal{T}$ along the $t$ axis and $\mathcal{L}$ along the spatial direction, then the (thermal) vacuum expectation value of $W(C)$ is given, in the limit $\lim_{\mathcal{T} \to \infty}$  by
\begin{equation}
\label{eq:wilsonrec}
\langle W(C) \rangle_0 \sim e^{i \mathcal{T} V_{Q\bar{Q}}(L,T)},
\end{equation}
where $V_{Q \bar{Q}}(L,T)$ is interpreted as the interaction energy of the $Q\bar{Q}$ pair and $T$ is the temperature of the thermal vacuum - for definiteness, we will call $V_{Q \bar{Q}}(L,T)$ the heavy quark-antiquark potential at finite temperature.

The interaction energy $V_{Q \bar{Q}}(L,T)$ may possess, at finite temperature, a finite imaginary part, which can be used to estimate a thermal width of the quarkonium, as shown in \cite{imvrefs,otherrefs1,otherrefsImV}. Calculations of $\mathrm{Im} V_{Q \bar{Q}}(L,T)$ relevant to QCD and heavy ion collisions were performed for static $Q\bar{Q}$ pairs using, for instance, perturbative QCD (pQCD) \cite{Laine:2006ns}, lattice QCD \cite{Rothkopf:2011db,Aarts:2011sm,Aarts:2013kaa} and using the gauge/gravity duality \cite{Noronha:2009da, Albacete:2009bp, Hayata:2012rw, Finazzo:2013rqy, Fadafan:2013bva,Fadafan:2013coa} (for recent studies about quantum decoherence effects in quarkonia see \cite{Akamatsu:2011se,Akamatsu:2014qsa}). However, all of these calculations of $\mathrm{Im} V_{Q \bar{Q}}(L,T)$ in the literature were performed considering static $Q\bar{Q}$ pairs. To accurately determine the suppression of quarkonia formed in heavy ion collisions it is necessary to evaluate $\mathrm{Im} V_{Q \bar{Q}}(L,T)$ for moving quarkonia in the QGP \cite{Strickland:2011mw,Strickland:2011aa,mikenew,Aarts:2012ka}. In Ref.\ \cite{Escobedo:2013tca}, it was shown using effective theory techniques that in a weakly coupled QGP the heavy quarkonia decay width is a nontrivial function of the temperature and velocity. However, for the two different scenarios considered in \cite{Escobedo:2013tca} involving the different scales in the problem, it was found that the decay width of very rapidly moving quarkonia decreases with the pair's velocity.

From the viewpoint of holography and the gauge/gravity duality \cite{Maldacena:1997re,Witten:1998qj,Witten:1998zw}, the evaluation of $\langle W(C) \rangle$ (and thus of $V_{Q \bar{Q}}(L,T)$) in the large $N_c$ strongly coupled 4-dimensional gauge theory corresponds to, in the 5-dimensional bulk geometry perspective of the gravity dual, the problem of finding a classical string configuration that has the closed loop $C$ as the boundary of the string worldsheet in the bulk \cite{Maldacena:1998im}. The seminal calculations for $T=0$ for strongly coupled $\mathcal{N} = 4$ Super Yang-Mills (SYM) \cite{Maldacena:1998im} were extended to finite temperature \cite{Brandhuber:1998bs,Rey:1998bq} and also to more general backgrounds \cite{Kinar:1998vq,Sonnenschein:1999if}. Holographic calculations of the real part of $V_{Q \bar{Q}}(L,T)$ in strongly coupled plasmas, for moving $Q\bar{Q}$ pairs, were already considered by Liu et al. \cite{Liu:2006nn,Liu:2006he} in the case of a strongly coupled $\mathcal{N} = 4$ SYM plasma. The computation of $\mathrm{Re} \, V_{Q \bar{Q}}(L,T)$ for moving $Q\bar{Q}$ in more general backgrounds dual to strongly coupled QFTs were studied in Ref.\ \cite{Caceres:2006ta}.

In Ref.\ \cite{Faulkner:2008qk}, the momentum dependence of meson widths was computed within the gauge/gravity duality and it was shown that this quantity receives nontrivial contributions from instantons on the string worldsheet. An interesting feature of their approach is that the thermal width becomes very large for rapidly moving mesons. Thus, while the general arguments from Refs.\ \cite{Liu:2006nn,Liu:2006he,Ejaz:2007hg} indicate that the dissociation temperature of mesons decreases with the pair's rapidity, the results of Ref.\ \cite{Faulkner:2008qk} show that even before complete dissociation the imaginary part of rapidly moving mesons may be already large enough to cause suppression of these states in a strongly coupled plasma. 

A general approach to determine the imaginary part of the static heavy quark potential using string worldsheet fluctuations was developed in \cite{Noronha:2009da, Finazzo:2013rqy}. In this paper, we generalize this method to estimate the imaginary part of $V_{Q \bar{Q}}(L,T)$ for moving quarkonia, starting from the evaluation of the real part as done in \cite{Liu:2006nn,Liu:2006he}. The main idea is, following \cite{Liu:2006nn,Liu:2006he}, to consider a boost from the frame where the plasma is at rest and the $Q\bar{Q}$ dipole is moving to a frame where the $Q\bar{Q}$ dipole is at rest, while the plasma is moving. The procedure presented in this work can be used for a large class of strongly coupled theories dual to gravity though in this work we focus on its application to strongly coupled $\mathcal{N} = 4$ SYM plasma. The method pursued here, however, has limitations. In the static case these limitations (that stem from the saddle point approximation used in the calculation of the imaginary part of the potential) were discussed in detail in \cite{Finazzo:2013rqy} and we shall see in this paper that similar limitations restrict our discussion here to the case of slowly moving quarkonia (therefore, our results are much more relevant to RHIC collisions than those at the LHC). We note, however, that the overall qualitative behavior found here, i.e., slowly moving quarkonia are less stable than the static case and that $Q\bar{Q}$ pairs are more stable when they are aligned with their velocity axis, is consistent with the findings of \cite{Liu:2006nn,Liu:2006he,Ejaz:2007hg,Faulkner:2008qk}.

This paper is organized as follows: in Section \ref{sec:perp}, we discuss the case where the $Q\bar{Q}$ dipole is moving perpendicularly to the axis that joins the $Q\bar{Q}$ pair - this presents a simpler problem where, in the case of strongly coupled $\mathcal{N} = 4$ SYM, some simple analytical results can be obtained. The case of a general orientation of the $Q\bar{Q}$ pair is a more complex problem that is dealt with in Section \ref{sec:arbang}, together with numerical calculations for this quantity in an $\mathcal{N} = 4$ SYM plasma. In Section \ref{sec:concl} we present our conclusions and outlook.

\section{Dipole Perpendicular To The Hot Wind}
\label{sec:perp}

\subsection{General results - Real part}

In this section we evaluate general expressions for the real and imaginary part of potential energy $V_{Q\bar{Q}}$ of quark-antiquark $Q\bar{Q}$ pair moving with the $Q\bar{Q}$ dipole axis oriented perpendicularly to a strongly coupled non-Abelian plasma, using holographic methods. This case is computationally simpler than the case of a dipole with an arbitrary orientation with respect to the wind. In fact, this case can be solved analytically in the case of strongly coupled $\mathcal{N}=4$ SYM and it represents an extreme case (and check) of the calculations of $V_{Q\bar{Q}}$ for arbitrary orientations of the dipole with respect to the wind. Our calculations for the real part follow the analysis done in \cite{Liu:2006nn,Liu:2006he}.

We start by assuming that our $d+1$-dimensional gauge theory in Minkowski space has a gravity dual with the following metric,
\begin{equation}
\label{eq:metricperp}
ds^2 = -G_{00} (U) dt^2 + G_{xx} (U) dx_i^2 + G_{UU} (U) dU^2,
\end{equation}
where $i=1,2,...,d$, $x_i$ are orthornormal spatial coordinates for the boundary and $U$ is the radial coordinate. We will assume that our gravity dual has an asymptotically $AdS_5$ boundary at $U \to \infty$ and a black brane horizon at $U=U_h$, where we will assume that $G_{UU} (U_h)\to \infty$, with $G_{00} (U_h) G_{UU} (U_h)$ finite. The presence of a black brane (which implies $G_{00} \neq G_{xx}$) breaks the original $SO(d,1)$ Lorentz isometry of the metric in the transverse spacetime coordinates $(t,x_i)$ to only a rotational $SO(d)$ isometry in the spatial coordinates $x_i$.

The fact that we do not have the full $SO(d,1)$ isometry group means that the metric \eqref{eq:metricperp} is not invariant under rigid Lorentz boosts of the Minkowski spacetime slices we have at each fixed $U$. This is expected, since the presence of a black brane in the gravity dual is associated with a thermal boundary field theory, and in this case there is a preferred reference frame (namely, the frame where the thermal medium is at rest). Boosting this frame  with a velocity $\vec{v}$ means that the an observer in this frame sees the medium moving past him with velocity $-\vec{v}$. 

We can exploit this fact to study the effect of the plasma on a $Q\bar{Q}$ pair in the thermal medium. Starting from a reference frame where the plasma is at rest and the $Q\bar{Q}$ dipole is moving with a constant velocity - we can boost to a reference frame where the dipole is at rest but the plasma is moving past it. This is the main idea used in this calculation. When interpreting the results of our calculations, we will interchange frequently between both points of view.

With these considerations in mind, let us consider a $Q\bar{Q}$ pair moving with rapidity $\eta$ along the $x_d$ direction with the plasma at rest in this reference frame. Let us then boost our reference frame in the $x_d$ direction with rapidity $\eta$, so that the $Q\bar{Q}$ is now at rest and the plasma moves with rapidity $-\eta$ in the $x_d$ direction (the $Q\bar{Q}$ now feels a hot wind):
\begin{align}
\label{eq:boost}
dt' = dt' \cosh \eta - dx'_d \sinh \eta \nonumber \\
dx_d = -dt' \sinh \eta + dx'_d \cosh \eta.
\end{align}
Applying this boost to the transverse coordinates of the metric \eqref{eq:metricperp}, the geometry now becomes (after dropping the primes):
\begin{align}
\label{eq:metricboost}
ds^2 = & -(G_{00} \cosh^2 \eta - G_{xx} \sinh^2 \eta) dt^2 + (G_{xx} \cosh^2 \eta - G_{00} \sinh^2 \eta) dx_d^2 + \nonumber \\ & -2(G_{xx}-G_{00}) \sinh \eta \,  \cosh \eta \, dt \, dx_d + G_{xx} dx_j^2 + G_{rr} dr^2,
\end{align}
where $j=1,2,..,d-1$.

Now, consider a $Q\bar{Q}$ dipole oriented perpendicularly to the wind in the gauge theory. Let $x_1$ be the direction to which the dipole is aligned and let $L$ be the length of the line joining both quarks. The quarks are located at $x_1 = L/2$ and $x_1 = -L/2$. As discussed in Section \ref{sec:intro}, the heavy quark-antiquark potential energy $V_{Q\bar{Q}}$ of this system is related to the expectation value of a rectangular Wilson loop by Eq.\ \eqref{eq:wilsonrec}. Holographically, in the supergravity limit (corresponding to a strongly coupled plasma) we can evaluate $\langle W(C) \rangle$ by the prescription \cite{Maldacena:1998im}
\begin{equation}
\label{eq:wilsonloohol}
\langle W(C) \rangle \sim e^{-i S_{str}}
\end{equation}
where $S_{str}$ is the classical Nambu-Goto action of a string in the bulk,
\begin{equation}
\label{eq:nambugoto}
S_{str} = -\frac{1}{2\pi \alpha'} \int d\sigma d\tau \sqrt{-det(G_{MN} \partial_{\alpha} X^M \partial_{\beta} X^N}),
\end{equation}
evaluated at an extremum of the action, $\delta S_{str} = 0$. The resulting equations of motion must be solved with the boundary condition that the worldsheet of the string, parametrized by spacetime target functions $X^M(\sigma,\tau)$ ($M=0,1,\ldots, d-1$ are the target spacetime indices, $\alpha,\beta =\sigma,\tau$  are the worldsheet coordinates), must describe the curve $C$, in the boundary of the bulk geometry. Plugging back $S_{str}$ in \eqref{eq:wilsonloohol} we extract the real part of $V_{Q\bar{Q}}$. Once we include thermal fluctuations of the string in \eqref{eq:wilsonloohol}, we will be able to evaluate the imaginary part of $V_{Q\bar{Q}}$ in this case.

Since the dipole is perpendicular to the wind, $x_{d-1}$ is constant along the line joining the endpoints of the string - this means that we can take $X^{d-1}$ to be constant. We use the remaining symmetry of \eqref{eq:nambugoto} to completely fix the static gauge given by $(X^0 = \tau = t, X^1 = \sigma = x, X^i=\mathrm{const}, X^{d-1} = \mathrm{const}, U = U(\sigma))$, where $i=1,...,d-2$. With this gauge choice, \eqref{eq:nambugoto} becomes, after inserting the background metric \eqref{eq:metricboost},
\begin{equation}
\label{eq:nambugotoperpstatic}
S_{str} = -\frac{\mathcal{T}}{2\pi \alpha'} \int_{-L/2}^{L/2} d\sigma \sqrt{\tilde{M}(U) U'(\sigma)^2 +  \tilde{V}(U)},
\end{equation}
where we defined
\begin{subequations}
\label{eq:MVtilde}
\begin{align}
\tilde{M} (U) \equiv M(U) \cosh^2 \eta - N(U) \sinh^2 \eta \\
\tilde{V} (U) \equiv V(U) \cosh^2 \eta - P(U) \sinh^2 \eta
\end{align}
\end{subequations}
and
\begin{subequations}
\label{eq:MVPNfuncdef}
\begin{align}
M(U) \equiv G_{00} G_{UU} \\
V(U) \equiv G_{00} G_{xx} \\
P(U) \equiv G_{xx}^2 \\
N(U) \equiv G_{xx} G_{rr}
\end{align}
\end{subequations}
Also, $U' \equiv dU/d\sigma$. We see that \eqref{eq:nambugotoperpstatic} has formally the same form found in the case of a plasma at rest - we only need to replace $\tilde{M}$ and $\tilde{V}$ by $M$,$V$, respectively \cite{Finazzo:2013rqy}. We also see that taking $\eta \to 0$ takes \eqref{eq:nambugotoperpstatic} back to the case in which the plasma is at rest. However, an important difference between \eqref{eq:nambugotoperpstatic} and the corresponding action in the case where the plasma is at rest is that in the latter the function $M$ is always positive whereas here $\tilde{M}$ can be negative, depending on the bulk geometry and $\eta$. 

Let us proceed to solve the variational problem $\delta S =0$. Since the calculation is very similar to the $\eta =0$ case (see, for example, \cite{Maldacena:1998im,Brandhuber:1998bs,Kinar:1998vq,Sonnenschein:1999if}) our notation closely follows \cite{Noronha:2009da,Finazzo:2013rqy}) and here we will only sketch the basic steps. First we write down the Hamiltonian associated with \eqref{eq:nambugotoperpstatic}, which is a constant of motion. Then, since the string is symmetric with respect to $X_1 = \sigma = 0$, we have $U'(0) = 0$, with the corresponding position of the deepest position in the bulk being $U(0) = U_c$. From the Hamiltonian, we can write the equation of motion for $U(\sigma)$,
\begin{equation}
\label{eq:eomperp}
\frac{dU}{d\sigma} = H(U(\sigma))
\end{equation}
with $H(U)$ defined by
\begin{equation}
\label{eq:Hdef}
H(U) \equiv \sqrt{\frac{\tilde{V}(U)}{\tilde{V}_c} \frac{\tilde{V}(U)-\tilde{V}_c}{\tilde{M}(U)}}\,.
\end{equation}
The subscript $c$ in $\tilde{V}_c$ and $\tilde{M}_c$ means that we evaluate these functions at $U=U_c$, i.e., $F_c \equiv F(U_c)$. It follows by the chain rule that
\begin{equation}
\label{eq:eomperp}
\frac{d^2U}{d\sigma^2} = \frac{1}{2} \frac{dH}{dU} (U(\sigma))
\end{equation}
and in particular
\begin{equation}
\frac{d^2U}{d\sigma^2} (\sigma = 0) = \frac{1}{2} \frac{\tilde{V}'_c}{\tilde{M}_c},
\label{eq:d2Uperp}
\end{equation}
where $\tilde{V}' \equiv d\tilde{V}/dU$. This equation will be useful later in the calculation of $\mathrm{Im} \, V_{Q\bar{Q}}$. From \eqref{eq:eomperp}, using the boundary condition that at the boundary of the bulk geometry the string has to reach the Wilson loop contour - or more precisely, $U(\sigma \to L/2) = \Lambda$, where $\Lambda$ is an UV cutoff, we can relate $L$ with $U_c$ as follows
\begin{equation}
\frac{L}{2} = \int\limits_{U_c}^{\Lambda} dr \frac{1}{\sqrt{H(U)}}\,.
\label{eq:Lperp}
\end{equation}

Plugging \eqref{eq:eomperp} back in \eqref{eq:nambugotoperpstatic} we can relate $S_{str}$ and $U_c$
\begin{equation}
\label{eq:Sperpnreg}
S_{str} = \frac{\mathcal{T}}{\pi \alpha'} \int\limits_{U_c}^{\Lambda} dU \, \sqrt{\tilde{M}(U)} \sqrt{\frac{\tilde{V}(U)}{\tilde{V}(U_*)}}  \left[\frac{\tilde{V}(U)}{\tilde{V}(U_*)}-1 \right]^{-1/2}.
\end{equation}
This action is formally infinite when we remove the cutoff UV $\Lambda$, since this means that the string worldsheet stretches from $U=U_c$ to the conformal boundary at $U\to \infty$ and thus has infinite area.  To regularize $S_{str}$, we subtract the $T \to 0$ divergence in $S_{str}$. This comes from the fact that in a thermal field theory all UV divergences should come from the vacuum. For further details, we refer to \cite{Finazzo:2013rqy}. The regularized Wilson loop is, therefore,
\begin{align}
\label{eq:Sperpreg}
S^{reg}_{str} & = \frac{\mathcal{T}}{\pi \alpha'} \left\{\int\limits_{U_c}^{\infty} dU \, \sqrt{\tilde{M}(U)} \sqrt{\frac{\tilde{V}(U)}{\tilde{V}(U_*)}}  \left[\frac{\tilde{V}(U)}{\tilde{V}(U_*)}-1 \right]^{-1/2} - \sqrt{M_0(U)} \right\} \nonumber \\ & - \frac{\mathcal{T}}{\pi \alpha'} \int_{U_h}^{U_c} \sqrt{M_0(U)},
\end{align}
where $M_0 (U) \equiv G^0_{00} (U) G^0_{UU} (U)$, with $G^0_{\mu \nu}$ being the metric in the absence of the black brane. Using \eqref{eq:Sperpreg} and \eqref{eq:wilsonloohol} we have, finally, $\mathrm{Re}\,V_{Q\bar{Q}} = S^{reg}_{str}/\mathcal{T}$ as a function of $U_c$. Together with \eqref{eq:Lperp} we can find $L(U_c)$ and $\mathrm{Re}\,V_{Q\bar{Q}} (U_c)$.

\subsection{Comments on the imaginary part}

The main idea to evaluate $\mathrm{Im} \, V_{Q\bar{Q}}$  is to consider thermal fluctuations of the string worldsheet \cite{Noronha:2009da,Finazzo:2013rqy}. The action \eqref{eq:nambugotoperpstatic} is exactly of the form already considered in Refs.\ \cite{Noronha:2009da,Finazzo:2013rqy}. However, now the function $\tilde{M}(U)$ is not strictly positive. When $\tilde{M}(U)>0$ the argument used in the calculation of the imaginary part from \cite{Noronha:2009da,Finazzo:2013rqy} is left unchanged. However, when $\tilde{M} (U)$ is negative, the thermal fluctuations that may generate an imaginary part must take place away from $x_1= 0$ and, thus, must have large amplitudes. Fluctuations of this kind cannot be considered using the current approximations employed in our approach, as we discuss below.

\begin{figure}[h!t]
\begin{center}
\includegraphics[width=0.7 \textwidth]{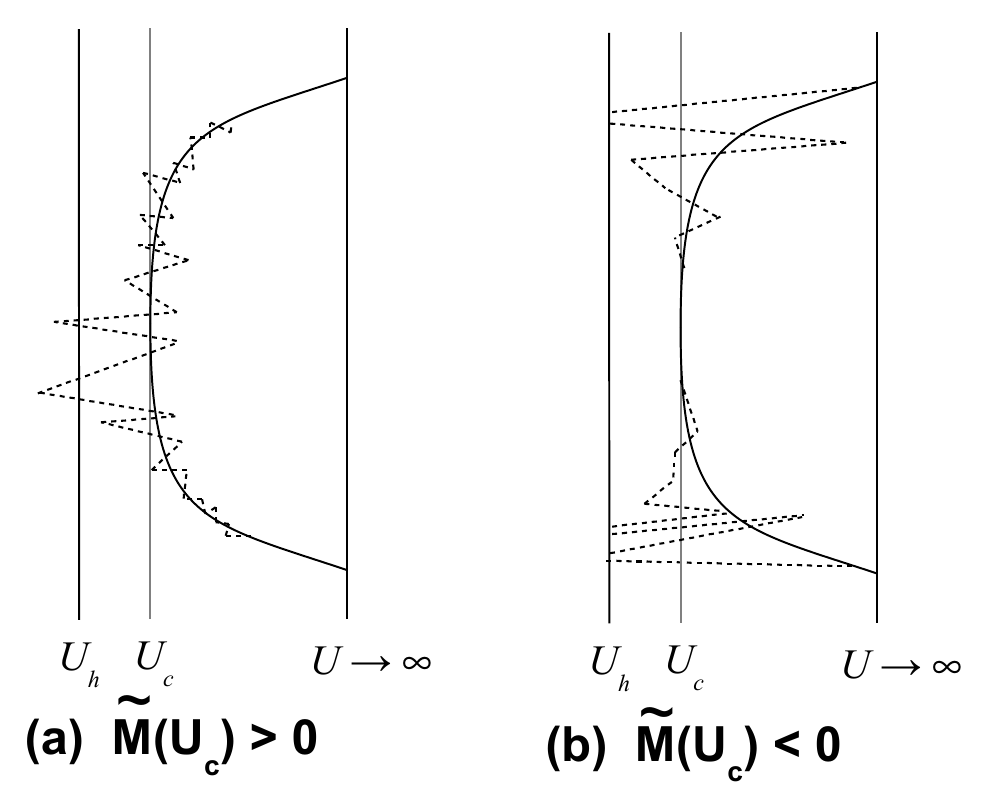}
\caption{Thermal fluctuations of the string worldsheet responsible for $\mathrm{Im} \, V_{Q\bar{Q}} \neq 0$ when (a) $\tilde{M}(U)>0$ and (b) $\tilde{M}(U)<0$. Fluctuations of the second type, large and distant from $U=U_c$ cannot be considered in our current approach since they require corrections that go beyond the saddle point approximation.}
\label{fig:thermalfluchotwinds}
\end{center}
\end{figure}

If $\tilde{M}(U_c) > 0$, the argument in \cite{Noronha:2009da,Finazzo:2013rqy} can be readily used. We refer the reader to \cite{Finazzo:2013rqy} for the necessary details. In the end, one can show that
\begin{equation}
\label{eq:ImFQQperp}
\mathrm{Im} \, V_{Q\bar{Q}} = -\frac{1}{2\sqrt{2} \alpha'} \sqrt{\tilde{M}_c} \left[ \frac{\tilde{V}'_c}{2 \tilde{V}''_c} - \frac{\tilde{V}_c}{\tilde{V}'_c} \right],
\end{equation}
if $\mathrm{Im} \, V_{Q\bar{Q}} <0$.

If, instead, $\tilde{M}(U_c) < 0$, the argument is the same up to the point where we write Eq. (4.6) in \cite{Finazzo:2013rqy}
\begin{equation}
\mathcal{L}_j = \sqrt{C_1 x_j^2 + C_2}
\label{eq:Ljperp}
\end{equation}
with $C_1$ and $C_2$ given by Eqs.\ (4.7) and (4.8) in \cite{Finazzo:2013rqy}, with the substitutions $M,V \to \tilde{M},\tilde{V}$. The argument of the square root in \eqref{eq:Ljperp} is, now, positive for a small interval of $x_j$ and negative outside of this interval. This means that the square root is purely real inside this interval and negative outside this interval. Thus, $\mathcal{L}_j$ has an imaginary part only outside of the interval centered in $x_1 = 0$. However, our calculation is only valid for $x_1 \sim 0$, near the bottom of the string - we do not have access to the fluctuations that may occur far from $x_1 =0$ since they would require to go beyond the saddle point approximation. Thus, there is still an imaginary part for the potential in this case but, due to the saddle point approximation, we cannot compute it with the formalism derived in \cite{Noronha:2009da,Finazzo:2013rqy}. This interesting case, which corresponds to the case of large $Q\bar{Q}$ rapidities, is beyond the range of applicability of the approximations employed in our method and a more general construction involving the explicit D-branes degrees of freedom \cite{Karch:2002sh} corresponding to fundamental quarks used in \cite{Faulkner:2008qk} may be required\footnote{We note that even though the worldsheet fluctuation method \cite{Noronha:2009da,Finazzo:2013rqy} is formally applicable in general, at the moment we have not worked out the technically nontrivial issues needed to go beyond the saddle point approximation.}.

\subsection{An explicit example - Thermal $\mathcal{N} = 4$ SYM}

Let us apply the results of the foregoing subsection for the case of $\mathcal{N} = 4$ SYM plasma. The gravity dual to thermal $\mathcal{N} = 4$ SYM is type IIB superstring theory on $AdS_5 \times S^5$ with a black brane. The metric of the gravity dual is
\begin{equation}
\label{eq:metricads}
ds^2 = -\frac{U^2}{R^2} f(U) dt^2 + \frac{U^2}{R^2} d\vec{x}^2 + \frac{R^2}{U^2} \frac{1}{f(U)} dU^2 + R^2 d\Omega_5^2.
\end{equation}
where $f(U) = 1- U_h^4/U^4$ is the blackening factor and $R$ is the common $AdS_5$ and $S^5$ radius. The temperature of the black brane and, therefore, of the thermal gauge theory, is $T=U_h/(\pi R^2)$. We choose a fixed configuration on the $S^5$ and, thus, we neglect its contribution to the dynamics. It follows that the functions in \eqref{eq:MVPNfuncdef} are
\begin{subequations}
\begin{align}
\label{eq:MVPNfuncAdS}
M(U) =1 \\
V(U) = \frac{U^4}{R^4} \left( 1- \frac{U_h^4}{U^4} \right) \\
P(U) = \frac{U^4}{R^4} \\
N(U) = \left(1- \frac{U_h^4}{U^4} \right)^{-1}\,.
\end{align}
\end{subequations}
After some algebra, one obtains from \eqref{eq:Lperp} and \eqref{eq:Sperpreg}
\begin{equation}
\label{eq:LAdSperp}
LT = \frac{2 y_c}{\pi} \sqrt{1- y_c^4 \cosh^2 \eta} \, \int_1^{\infty} \frac{dy}{\sqrt{(y^4-1)(y^4-y_c^4)}}\,,
\end{equation}
\begin{equation}
\label{eq:FQQAdSperp}
\frac{\mathrm{Re}\,V_{Q\bar{Q}}}{T} =\frac{\sqrt{\lambda}}{y_c} \left\{ \int_1^{\infty} dy \, \left[\frac{y^4}{\sqrt{(y^4-1)(y^4-y_c^4)}} - 1 \right] - \frac{1}{y_h} \right\},
\end{equation}
where $y_h \equiv U_h/U_c$, $y_c \equiv = 1/y_h$ and $\lambda = R^4/\alpha'^2$ is the 't Hooft coupling of the gauge theory. Note that since $U_c > U_h$ one has $0<y_c<1$. It is possible to integrate \eqref{eq:LAdSperp} and \eqref{eq:FQQAdSperp} analytically (see the Appendix B of Ref. \cite{Finazzo:2013rqy} for details; the main idea is to use integral representations of the hypergeometric function \cite{gradshteyn})
\begin{equation}
\label{eq:LAdSperp}
LT = \frac{2 \sqrt{2 \pi} y_c}{\Gamma(1/4)^2} \sqrt{1- y_c^4 \cosh^2 \eta} \, {}_2 F_1 \left(\frac{1}{2}, \frac{3}{4}, \frac{5}{4}, y_c^4 \right)\,,
\end{equation}
\begin{equation}
\label{eq:FQQAdSperp}
\frac{\mathrm{Re}\,V_{Q\bar{Q}}}{T} = - \frac{\sqrt{\lambda}}{y_c} \frac{\sqrt{2 \pi^3}}{\Gamma(1/4)^2} \, \left[ {}_2 F_1 \left(\frac{1}{2}, -\frac{1}{4}, \frac{1}{4}, y_c \right) + y_c^4 \cosh^2 \eta \, {}_2 F_1 \left(\frac{1}{2}, \frac{3}{4}, \frac{5}{4}, y_c^4 \right)  \right].
\end{equation}
where ${}_2 F_1(a,b,c,d)$ is the Gauss hypergeometric function. The limit $\eta \to 0$ yields the expressions shown in \cite{Finazzo:2013rqy}; for \eqref{eq:LAdSperp} this is immediate, whereas for \eqref{eq:FQQAdSperp} it requires some use of the properties of hypergeometric functions (namely, Gauss recursion formulas) \cite{gradshteyn}.

\begin{figure}[h!t]
\begin{center}
\includegraphics[width=0.6 \textwidth]{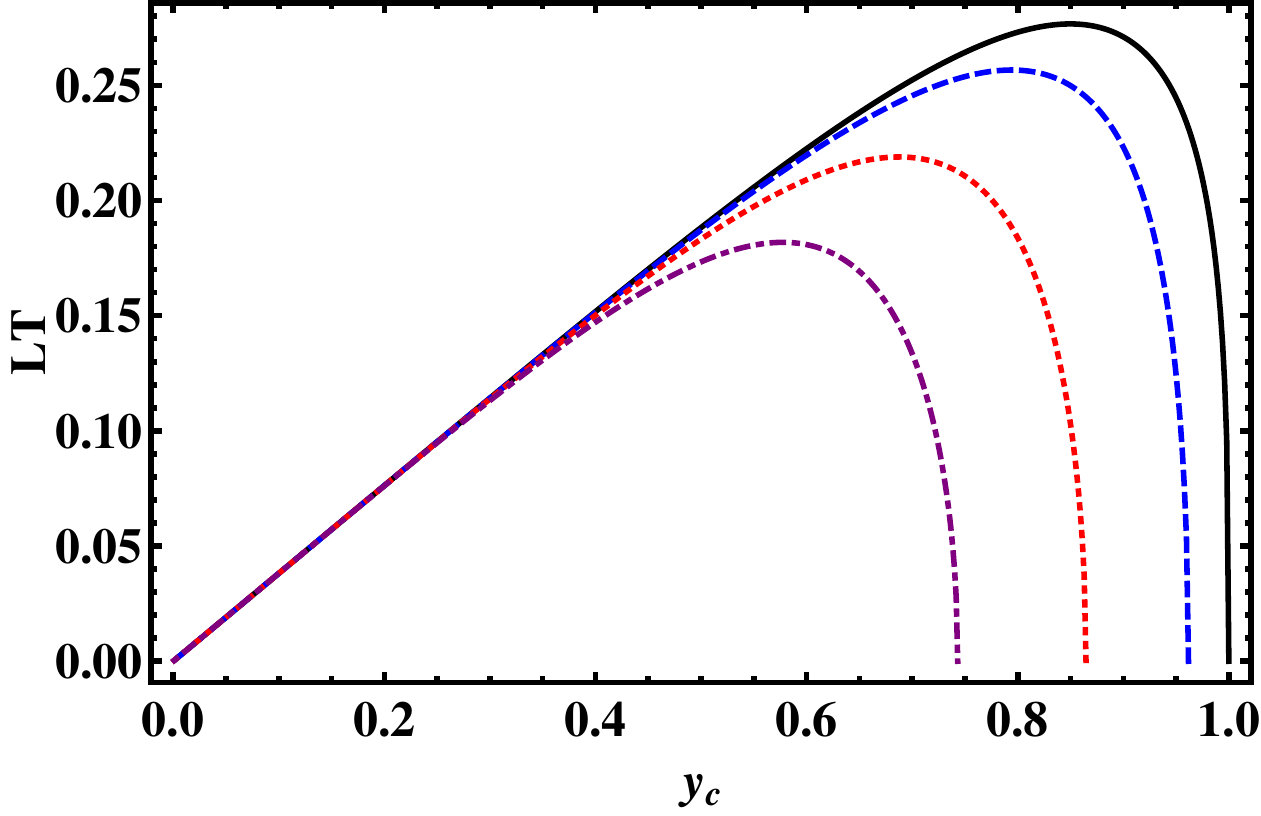}
\caption{(Color online) LT as a function of $y_c$ for a $Q\bar{Q}$ pair oriented perpendicularly to the hot wind in an $\mathcal{N}=4$ SYM plasma. Different rapidities are considered: the solid black curve corresponds to $\eta = 0$, the dashed blue curve to $\eta=0.4$, the dotted red curve to $\eta=0.8$, and the dashed-dotted purple curve to $\eta = 1.2$.}
\label{fig:LTycfunc}
\end{center}
\end{figure}

\begin{figure}[h!t]
\begin{center}
\includegraphics[width=0.6 \textwidth]{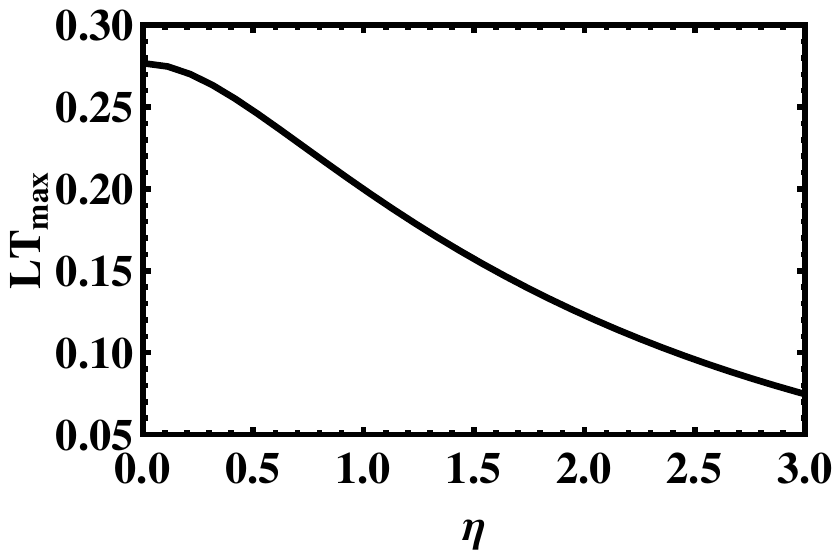}
\caption{$LT_{max}$ as a function of rapidity $\eta$ for a $Q\bar{Q}$ pair oriented perpendicularly to the hot wind in an $\mathcal{N}=4$ SYM plasma.}
\label{fig:LTmaxeta}
\end{center}
\end{figure}

In Fig.\ \ref{fig:LTycfunc} we show the behavior of $LT$ for this perpendicular case as a function of $y_c$ for several choices of $\eta$. The maximum of the $LT(y_c)$, $LT_{max}$ indicates the limit of validity of the saddle point approximation - to go to higher $LT$ it is necessary to include further connected contributions past the saddle point approximation \cite{Bak:2007fk}. We see from Fig.\ \ref{fig:LTmaxeta} that increasing $\eta$ reduces $LT_{max}$. A systematic study of $LT_{max}$ as a function of $\eta$ is presented in Fig.\ \ref{fig:LTmaxeta}.
In the literature \cite{Liu:2006nn,Liu:2006he}, $LT_{max}$ has been used to define a dissociation length for the moving $Q \bar{Q}$ pair - the dominant configuration for $S_{str}$ in this case would be two straight strings (two heavy quarks) running from the boundary to the horizon. In this paper, based on the discussion in \cite{Bak:2007fk,Finazzo:2013rqy}, we choose to use this quantity to define the region of applicability of the U-shaped string configuration (which is dependent on the pair's rapidity). Moreover, the dissociation properties of heavy quarkonia should be sensitive to the imaginary part of the potential and that will be estimated later in this section using the expression for $\mathrm{Im} \, V_{Q\bar{Q}}$ computed above.

\begin{figure}[h!t]
\begin{center}
\includegraphics[width=0.6 \textwidth]{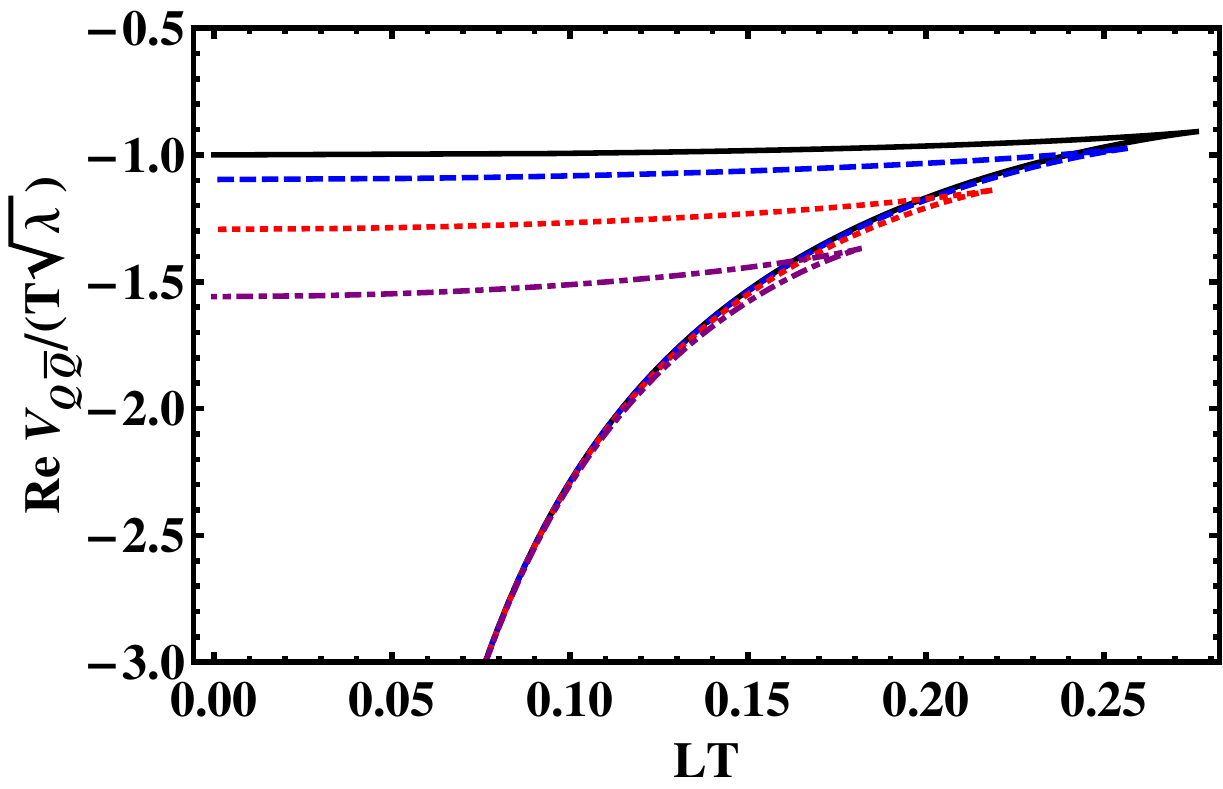}
\caption{(Color online) $\mathrm{Re}\,V_{Q\bar{Q}}/(T\sqrt{\lambda})$ as a function of $LT$ for a $Q\bar{Q}$ pair oriented perpendicularly to the hot wind in an $\mathcal{N}=4$ SYM plasma. Different rapidities are considered: the solid black curve corresponds to $\eta = 0$, the dashed blue curve to $\eta=0.4$, the dotted red curve to $\eta=0.8$ and the dashed-dotted purple curve to $\eta = 1.2$.}
\label{fig:FQQLTfunc}
\end{center}
\end{figure}

We proceed to show, in Fig.\ \ref{fig:FQQLTfunc}, $\mathrm{Re} \,V_{Q\bar{Q}}/(T\sqrt{\lambda})$ as a function of $LT$ for some choices of $\eta$. We see that for short distances the $Q\bar{Q}$ pair does not feel the moving plasma, as expected. For each $\eta$ the upper branch corresponds to another saddle point of the string action (which is associated to the curve to the right of $LT_{max}$ in Fig.\ \ref{fig:LTycfunc}), which is suppressed with respect to the lower branch.

\begin{figure}[h!t]
\begin{center}
\includegraphics[width=0.6 \textwidth]{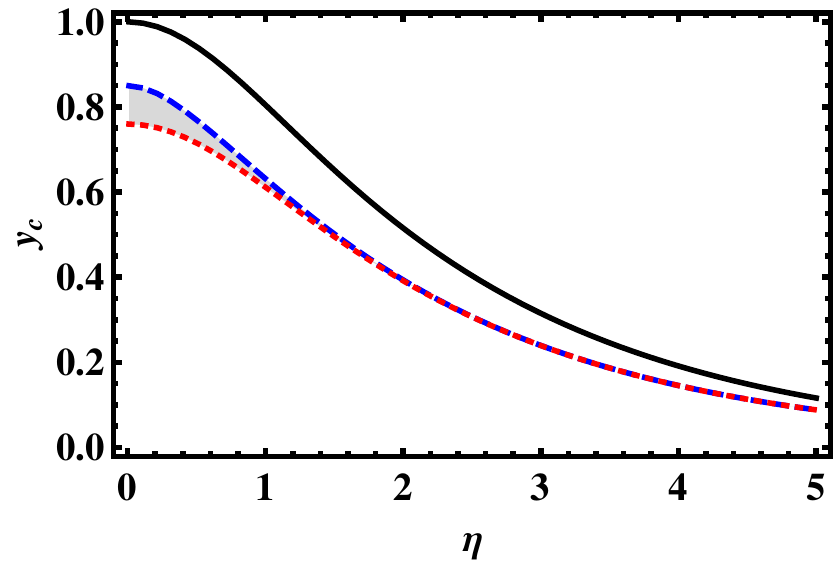}
\caption{(Color online) From top to bottom, the limiting curves $y_{max,1}$ (solid black line), $y_{max,2}$ (dashed blue line) and $y_{min}$ (dotted red curve) as a function of the rapidity $\eta$ for a $Q\bar{Q}$ pair oriented perpendicularly to the hot wind in an $\mathcal{N}=4$ SYM plasma. The filled area represents the region where our method can be reliably used to estimate $\mathrm{Im}\,V_{Q\bar{Q}}$.}
\label{fig:allowed}
\end{center}
\end{figure}

Now, let us proceed to the evaluation of the imaginary part in this case. First, the condition $\tilde{M}(U_c) > 0$ leads to \begin{equation}
\label{eq:adsperpcond2}
y_c < y_{max,1} = (1-\tanh^2 \eta)^{1/4}\,.
\end{equation}
When this is valid, use of \eqref{eq:ImFQQperp} yields, after some algebra,
\begin{equation}
\label{eq:ImFQQperpads}
\frac{\mathrm{Im}\,V_{Q\bar{Q}}}{T} = -\frac{\pi}{24\sqrt{2}} \frac{\sqrt{\lambda}}{y_c} \sqrt{\frac{1-y_c^4 \cosh^2\eta}{1-y_c^4}} (3 y_c^4 \cosh^2 \eta-1)\,.
\end{equation}
Imposing that $\mathrm{Im}\,V_{Q\bar{Q}} < 0$ leads to
\begin{equation}
\label{eq:adsperpcond2}
y_c > y_{min} = \frac{1}{3^{1/4} \sqrt{\cosh^2\eta}}\,.
\end{equation}
Also, we must take $y_c < y_{max,2}$, where $y_{max,2}$ is the maximum value of $y_c$ for which the connected contribution we consider is valid (see Fig.\ \ref{fig:LTycfunc}). These conditions lead to a narrow window where our method is applicable, shown in Fig.\ \ref{fig:allowed}. One sees that $y_{max,2} < y_{max,1}$ so the case $\tilde{M}(U_c) <0$ does not need to be considered in our case. For $y < y_{min}$, $\mathrm{Im}\,V_{Q\bar{Q}} = 0$. For $y_{max,2}<y<y_{max,1}$, our method is not applicable and the imaginary part of the potential has to be computed using other methods. 

\begin{figure}[h!t]
\begin{center}
\includegraphics[width=0.6 \textwidth]{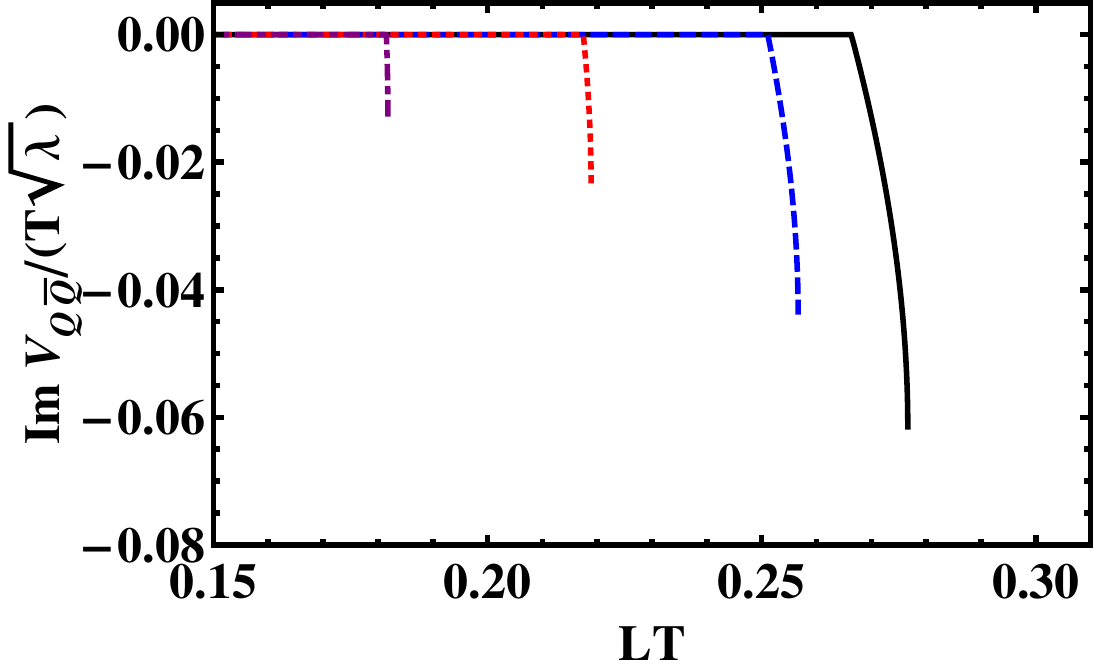}
\caption{(Color online) $\mathrm{Im}\,V_{Q\bar{Q}}/(\sqrt{\lambda}T)$ as a function of $LT$ for a $Q\bar{Q}$ pair oriented perpendicularly to the hot wind in an $\mathcal{N}=4$ SYM plasma. The solid black curve corresponds to $\eta = 0$, the dashed blue curve to $\eta=0.4$, the dotted red curve to $\eta=0.8$, and the dashed-dotted purple curve to $\eta = 1.2$.}
\label{fig:ImFQQLTfunc}
\end{center}
\end{figure}

Taking into account these intervals of applicability, we show the result of \eqref{eq:ImFQQperpads} for $\mathrm{Im}\,V_{Q\bar{Q}}$ as a function of $LT$ in Fig.\ \ref{fig:ImFQQLTfunc} for some choices of the rapidity $\eta$. One can see, from \eqref{eq:ImFQQperpads}, that $\mathrm{Im}\,V_{Q\bar{Q}}$ roughly scales as $T^2$ (since $y_c = U_h/U_c \propto T$). This scaling was seen in a calculation of $\mathrm{Im}\,V_{Q\bar{Q}}$ using complex world-sheet coordinates \cite{Albacete:2009bp} and in lattice calculations \cite{Rothkopf:2011db}, opposed to the $T$ scaling predicted by pQCD \cite{Laine:2006ns}. Also, for increasing rapidity, the onset of the imaginary part happens for smaller $LT$ which indicates that the suppression becomes stronger. However, the actual magnitude of the imaginary part computed in this case must be interpreted with caution. The apparent smaller magnitude observed at larger rapidities happens not because the imaginary part (or, equivalently, the thermal width) really decreases but simply because we chose to plot only the values that are consistent with the very stringent requirements used in \cite{Finazzo:2013rqy}. One could have used the linear extrapolation employed in the original study \cite{Noronha:2009da} and that would give the correct qualitative result that at large rapidity $Q\bar{Q}$ pairs are less stable. However, in \cite{Finazzo:2013rqy} this extrapolation was shown to largely overestimate the thermal width and, thus, in this paper we chose to be very ``conservative" and plot in Fig.\ \ref{fig:ImFQQLTfunc} only the region consistent with the approximations used in our method. However, we stress that the correct way to interpret our findings is that the heavy quark potential of moving quarkonia should have, in general, larger imaginary parts than the corresponding static case.

\section{Dipole at arbitrary angles}
\label{sec:arbang}

The next step is to generalize the previous calculations for a dipole oriented at an arbitrary angle with respected to its velocity vector. The main difference, in the gravity dual calculation of the Wilson loop, is that we cannot take $X_{d-1} = const$ anymore. As emphasized in \cite{Liu:2006nn,Liu:2006he}, the system has now two effective degrees of freedom. The calculation of the imaginary part, though it follows the same general lines as before, now needs two pieces of information that can be extracted from the classical solution.

\subsection{General Results - Real Part}

\begin{figure}[h]
\begin{center}
\includegraphics[width=0.9 \textwidth]{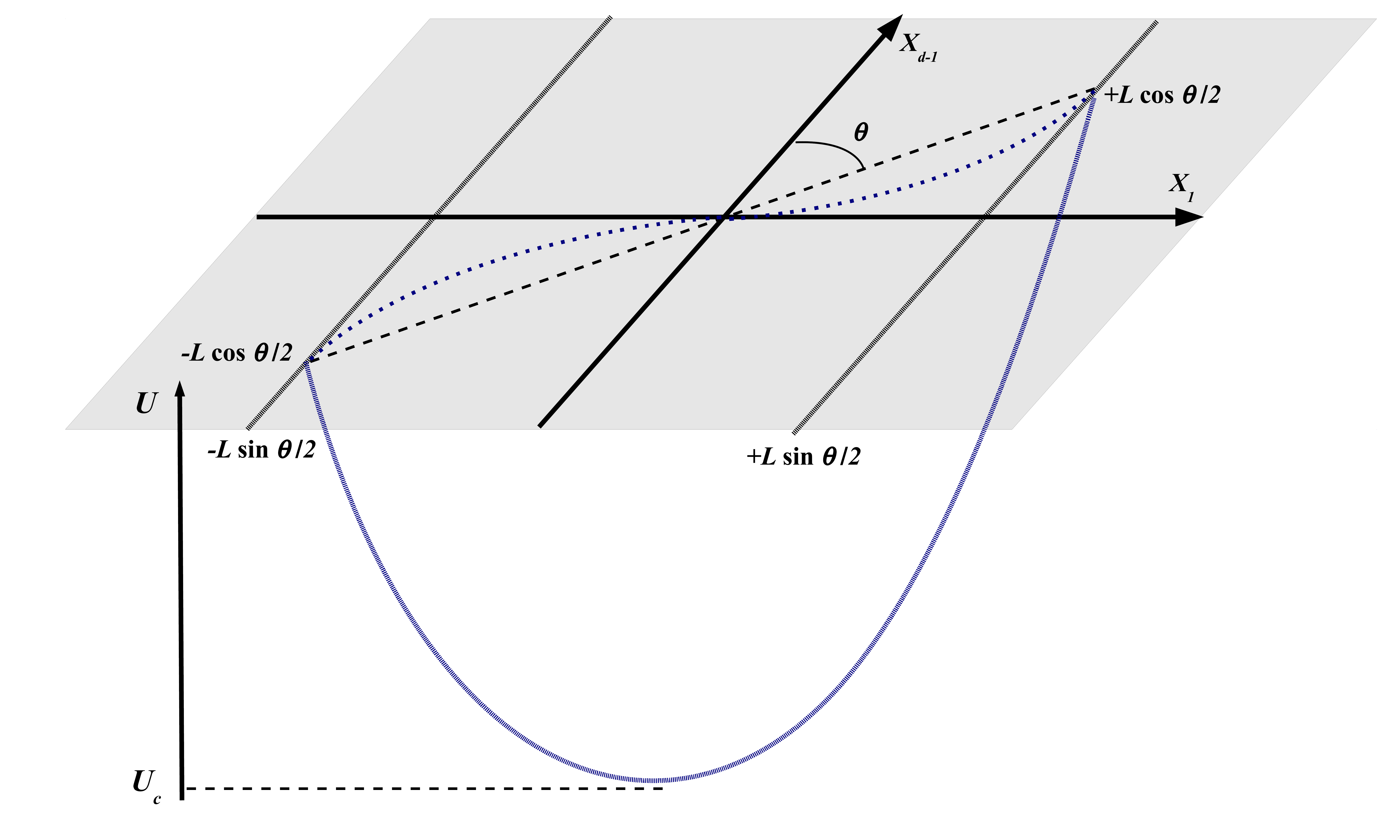}
\caption{(Color online) The geometry of the $\bar{Q}Q$ dipole, as described in the text, along with the string in the bulk joining the ends of the heavy quark pair. Note that the projection of this string on $(X_1, X_{d-1})$ at the boundary is not a straight line joining the $\bar{Q}Q$ quarks but a curved line.}
\label{fig:stringangle}
\end{center}
\end{figure}

Let us proceed to orient the $\bar{Q}Q$ dipole at an arbitrary angle with respect to the hot wind in the $X_{d-1}$ direction. Our objective will be, as before, to extract the real and imaginary parts of the heavy quark-antiquark potential energy $V_{\bar{Q}Q}$ holographically. We will take the dipole to be in the $(X_1, X_{d-1})$ plane. Let $\theta$ be the angle of the dipole with respect to the $X_{d-1}$ axis (see Fig.\ \ref{fig:stringangle}). If $\theta = \pi/2$ the dipole is oriented along $X_1$ and we return to the case of the previous section.

Using the holographic prescription, in order to evaluate $V_{Q\bar{Q}}$ we should evaluate $S_{str}$ given by \eqref{eq:nambugoto} with the boundary condition that the string described by $S_{str}$ joins the endpoints of the dipole. However, now we have two fixed parameters (aside from $\eta$, which tells us how to rotate the Minkowski slices): the length of the dipole $L$ and the angle $\theta$. If we followed the calculations of the previous section of this paper, we would have only one constant of motion, given by the Hamiltonian of the problem, to relate to both $L$ and $\theta$. Indeed, as discussed in \cite{Liu:2006nn,Liu:2006he}, one has to consider also a ``sagging" of the string along the line segment joining the quark-antiquark pair, as described in Fig.\ \ref{fig:stringangle}. In terms of the string worldsheet, this means that have to consider the behavior of the embedding function $X_{d-1}(\sigma)$. We cannot, as in the previous section, take $X_{d-1}(\sigma)$ as given. Rather, $X_{d-1}(\sigma)$ must be taken as the second degree of freedom of the string worldsheet and that makes a difference both in the computation of the real and imaginary parts of the potential.

This implies that we must slightly modify the static gauge used in the previous section. A possible choice of the string worldsheet coordinates, which will be used in this work, is  $(X^0 = \tau = t, X^1 = \sigma = x, X^i=\mathrm{const}, X^{d-1} = X^{d-1} (\sigma), U = U(\sigma))$, where $i=1,...,d-2$. Now we have two degrees of freedom, $X_{d-1} (\sigma)$ and $U(\sigma)$. We note that if the projection of the string on the $(X_1, X_{d-1})$ plane was a straight line, then $X_{d-1} (\sigma)$ would be given by $\sigma/\tan \theta$. The boundary conditions to be imposed here are
\begin{align}
\label{eq:boundangle}
U\left(\pm \frac{L}{2} \sin \theta \right) & = \Lambda \nonumber \\
X_d \left(\pm \frac{L}{2} \sin \theta \right) & = \pm \frac{L}{2} \cos \theta
\end{align}
which, taken together, imply that the string has as its endpoints the heavy quarks. As in Section \ref{sec:perp}, $\Lambda$ is an UV cutoff. The background metric is still given by the generic boosted metric in \eqref{eq:metricboost}.

With these modifications, the action \eqref{eq:nambugoto} now becomes
\begin{equation}
\label{eq:nambugotoangle}
S_{str} = -\frac{\mathcal{T}}{2\pi \alpha'} \int d \sigma \mathcal{L},
\end{equation}
where we defined the Lagrangian
\begin{equation}
\label{eq:lagrangean}
\mathcal{L} \equiv \sqrt{\left[ M(U) \cosh^2 \eta - N(U)  \sinh^2 \eta \right] U'(\sigma)^2 + V(U) X_d'(\sigma)^2 + \left[ V(U) \cosh^2 \eta - P(U)  \sinh^2 \eta \right] }
\end{equation}
with the functions $M,N,V$ and $P$ defined as in \eqref{eq:MVPNfuncdef}. This action does not depend on $\sigma$ explicitly, so the associated Hamiltonian
\begin{equation}
\label{eq:hamiltonian}
\mathcal{H} \equiv Q \equiv \mathcal{L} - \frac{dU}{d\sigma} \frac{\partial \mathcal{L}}{\partial U'} - \frac{dX_d}{d\sigma} \frac{\partial \mathcal{L}}{\partial X_d'}
\end{equation}
is a constant of motion. Another constant of motion is
\begin{equation} 
\label{eq:momentaXd}
K \equiv \frac{\partial \mathcal{L}}{\partial X_d'},
\end{equation}
since $\mathcal{L}$ does not depend on $X_d$ explicitly. Inserting the Lagrangian \eqref{eq:lagrangean} in the first integrals \eqref{eq:momentaXd} and \eqref{eq:hamiltonian} we obtain, after some algebra,
\begin{align}
\label{eq:eqmotionA}
Q^2 \left[ M(U) \cosh^2 \eta - N(U)  \sinh^2 \eta \right] U'(\sigma)^2 + Q^2 V(U) X_{d-1}'(\sigma)^2 + \nonumber \\ + \left[ V(U) \cosh^2 \eta - P(U)  \sinh^2 \eta \right] \left\{Q^2- \left[ V(U) \cosh^2 \eta - P(U)  \sinh^2 \eta \right] \right\} = 0
\end{align}
and
\begin{align}
\label{eq:eqmotionB}
K^2 \left[ M(U) \cosh^2 \eta - N(U)  \sinh^2 \eta \right] U'(\sigma)^2 + V(U) (K^2-V(U)) \, X_{d-1}'^2(\sigma) + \nonumber \\ + K^2 \left[ V(U) \cosh^2 \eta - P(U)  \sinh^2 \eta \right]=0\,.
\end{align}
Eqs.\ \eqref{eq:eqmotionA} and \eqref{eq:eqmotionB} can be recast in a more convenient form; first we solve \eqref{eq:eqmotionA} for $X_d'^2$, then we substitute the resulting expression in \eqref{eq:eqmotionB} to obtain an equation involving only $U'(\sigma)$. Then one can use the common term $\left[ M(U) \cosh^2 \eta - N(U)  \sinh^2 \eta \right] U'(\sigma)^2$ to obtain a simplified equation for $X_d'$. After these manipulations, one ends up with the final form for the equations of motion, namely
\begin{align}
\label{eq:eqmotionAfin}
Q^2 V(U) \left[ M(U) \cosh^2 \eta - N(U)  \sinh^2 \eta \right] U'(\sigma)^2 = \nonumber \\ = (V(U)-K^2)  \left[ V(U) \cosh^2 \eta - P(U)  \sinh^2 \eta \right]^2 - V(U) \left[ V(U) \cosh^2 \eta - P(U)  \sinh^2 \eta \right] Q^2
\end{align}
and
\begin{equation}
\label{eq:eqmotionBfin}
Q^2 V^2 (X_{d-1}')^2 = K^2 \left[ V(U) \cosh^2 \eta - P(U)  \sinh^2 \eta \right]^2.
\end{equation}

Referring back to the Fig.\ \ref{fig:stringangle}, we define $U_c$ as the minimum value of $U$ that the string reaches in the bulk. By the symmetry of the string worldsheet, we must have $U(\sigma=0) = U_c$, $U'(\sigma=0) = 0$ and $X_d (\sigma = 0) = 0$. Using these conditions in \eqref{eq:eqmotionAfin} we obtain an equation relating $U_c$ with $Q$ and $K$
\begin{equation}
\label{eq:relationUc}
(V_c -K^2) (V_c \cosh^2 \eta - P_c \sinh^2 \eta) - V_c Q^2 = 0,
\end{equation}
where we used, as before, $F_c \equiv F(U_c)$.

From the equations of motion \eqref{eq:eqmotionAfin} and \eqref{eq:eqmotionBfin}, and taking into account the boundary conditions \eqref{eq:boundangle}, we are lead to two relations between $L$, $\theta$ and $Q$,$K$ and $U_c$
\begin{align}
\label{eq:Lang1}
\frac{L}{2} \sin \theta = & Q \int_{U_c}^{\Lambda}\, dU \left\{ \frac{V(U)}{V(U) \cosh^2 \eta - P(U)  \sinh^2 \eta} \right. \times \nonumber \\ & \times \left. \frac{ M(U) \cosh^2 \eta - N(U)  \sinh^2 \eta}{\left[(V(U)-K^2)\left[V(U) \cosh^2 \eta - P(U)  \sinh^2 \eta\right] - V(U) Q^2 \right]} \right\}^{-1/2}
\end{align}
and
\begin{align}
\label{eq:Lang2}
\frac{L}{2} \cos \theta = K \int_{U_c}^{\Lambda}\, dU \, \sqrt{\frac{\left[M(U) \cosh^2 \eta - N(U)  \sinh^2 \eta \right] \left[V(U) \cosh^2 \eta - P(U)  \sinh^2 \eta\right]}{V(U)\left\{(V(U)-K^2)\left[V(U) \cosh^2 \eta - P(U)  \sinh^2 \eta\right] - V(U) Q^2 \right\}}}.
\end{align}
In these equations, if the metric is asymptotically AdS, the integrals converge and we can formally take $\Lambda \to \infty$. The general procedure, to be described in more details when we treat the explicit example of $\mathcal{N} =4$ SYM, is to specify a value for the constant $Q$ and solve \eqref{eq:relationUc} to obtain $U_c$ as a function of $Q$ and $K$. Then, plugging $U_c =U_c(Q,K)$, we may solve \eqref{eq:Lang1} and \eqref{eq:Lang2} to relate the set of integration constants $(Q,P)$ with the geometrical parameters $(L,\theta)$.

With the problem of the integration constants taken care of, we can evaluate the action in the saddle-point approximation and obtain
\begin{equation}
\label{eq:actionangnreg}
S = \frac{\mathcal{T}}{\pi \alpha'} \int_{U_c}^{\Lambda} \, dU \, \sqrt{\frac{ V(U) \left[M(U) \cosh^2 \eta - N(U)  \sinh^2 \eta \right] \left[V(U) \cosh^2 \eta - P(U)  \sinh^2 \eta\right]}{ \left\{(V(U)-K^2)\left[V(U) \cosh^2 \eta - P(U)  \sinh^2 \eta\right] - V(U) Q^2 \right\}}}\,.
\end{equation}
We regularize this action as before (the process of reorienting the string clearly cannot introduce new UV divergences). The regularized action is thus
\begin{align}
\label{eq:actionangreg}
S_{reg} & = \frac{\mathcal{T}}{\pi \alpha'} \int_{U_c}^{\Lambda} \, dU \, \left\{ \sqrt{\frac{ V(U) \left[M(U) \cosh^2 \eta - N(U)  \sinh^2 \eta \right] \left[V(U) \cosh^2 \eta - P(U)  \sinh^2 \eta\right]}{ \left\{(V(U)-K^2)\left[V(U) \cosh^2 \eta - P(U)  \sinh^2 \eta\right] - V(U) Q^2 \right\}}} \right. + \nonumber \\ & \left. - \sqrt{M_0 (U)} \right\} -\frac{\mathcal{T}}{\pi \alpha'} \int_0^{U_c} dU \sqrt{M_0 (U)},
\end{align}
with $M_0$ being, as before, the function $M(U)$ defined for $T=0$, in the absence of the black brane. The real part of the heavy quark potential of the $Q\bar{Q}$ pair is $\mathrm{Re}\, V_{Q\bar{Q}}=S_{reg}/\mathcal{T}$.

\subsection{Imaginary part - General results}

Let us proceed to discuss the imaginary part. As before, we consider thermal fluctuations in the string worldsheet, following the ideas in \cite{Noronha:2009da,Finazzo:2013rqy}. The main difference is that now we have two degrees of freedom, $U(\sigma)$ and $X_{d-1} (\sigma)$. Thus, in principle, we have to consider fluctuations $\delta U(\sigma)$ and $\delta X_{d-1}(\sigma)$, with $\partial U/\partial \sigma \to 0$ and $\partial X_{d-1}/\partial \sigma \to 0$. Including these fluctuations, the stringy partition function takes the form
\begin{equation}
\label{eq:thermalflucpartang}
Z_{str} \sim \int D(\delta U) \, D(\delta X_{d-1}) e^{i S_{str} (\bar{U}+\delta U, \bar{X}_{d-1} + \delta X_d)}
\end{equation}
where $\bar{U}$ and $\bar{X}_{d-1}$ are the classical solutions of $\delta S_{str} = 0$ described in the previous section. Partitioning the interval $[-L/2 \sin \theta , L/2 \sin \theta]$ in $2N$ subintervals and using the action \eqref{eq:nambugotoangle}, we can write
\begin{equation}
\label{eq:thermalflucpartang2}
Z_{str} \sim \left( \int_{-\infty}^{\infty} d(\delta U_{-N}) \, d(\delta X_{{d-1},-N}) \right) \cdots \left( \int_{-\infty}^{\infty} d(\delta U_{N}) \, d(\delta X_{{d-1},N}) \right) e^{i \frac{\mathcal{T} \Delta x}{2\pi \alpha'} \mathcal{L}_j},
\end{equation}
where $\Delta x = (L/2 \sin \theta)/2N$ and
\begin{equation}
\label{eq:lagrangeanthermalflucang}
\mathcal{L}_j = \sqrt{\tilde{M}(U(x_j)) (U'(x_j))^2 + V(U(x_j)) (X_{d-1}'(x_j))^2 + \tilde{V} (x_j)},
\end{equation}
with $\tilde{M}$ and $\tilde{V}$ given in \eqref{eq:MVtilde}. Only considering fluctuations near the bottom of the string, i.e., at $\sigma =0$ with $U = U_c$, we can expand the classical solution $\bar{U}(\sigma)$ around $\sigma =0$ to quadratic order on $\sigma$, and the functions $\tilde{M}$ and $\tilde{V}$ around $\bar{U}$ - the calculations are the same as for the static or perpendicular cases, noting that we keep only terms up to quadratic order in the monomial $\delta U^m \, \delta X_{d-1}^n \sigma^p$. As for $X_{d-1}(x_j)$, if the string did not sag, then we would have $X_{d-1} (\sigma) = \sigma/\tan \theta$. With the sagging, $X_{d-1}$ does not assume such a simple form. However, around $\sigma = 0$ one must be able to expand $X_{d-1} (\sigma)$ as
\begin{equation}
\label{eq:Xdexpansion}
X_{d-1}(\sigma) = \frac{\sigma}{\tan \tilde{\theta}} + b \sigma^3 + O (\sigma^5),
\end{equation}
where $\tilde{\theta}$ would be equal to $\theta$ if the string did not sag and $b$ was a constant. Even terms do not participate in this expansion because the problem is evidently symmetric under reflections with respect to the origin of the $(X_1,X_d)$ plane (see Fig.\ \ref{fig:stringangle} where the projection of the string was drawn taking this into account) and $X_d(\sigma)$ must be an odd function of $\sigma$. Therefore, up to $\mathcal{O}(\sigma^2)$,
\begin{equation}
\label{eq:Xdquad}
X_{d-1}'(\sigma)^2 = \frac{1}{\tan^2 \tilde{\theta}} + \frac{6 b}{\tan \tilde{\theta}} \sigma^2.
\end{equation}
Inserting \eqref{eq:Xdquad} into \eqref{eq:lagrangeanthermalflucang} we arrive, after some algebra, at
\begin{equation}
\label{eq:lagrangeanimang}
\mathcal{L}_j = \sqrt{\tilde{C}_1 x_j^2 + \tilde{C}_2},
\end{equation}
where we defined
\begin{equation}
\label{eq:C1tilde}
\tilde{C}_1 \equiv \tilde{M}_c \bar{U}''(0)^2+ \frac{1}{2} \left(\frac{V'_c}{\tan^2 \tilde{\theta}}+\tilde{V}'_c\right)\bar{U}''(0)+\frac{6 b}{\tan \tilde{\theta}} V_c
\end{equation}
and
\begin{equation}
\label{eq:C1tilde}
\tilde{C}_2 \equiv  \left(\frac{V_c}{\tan^2 \tilde{\theta}}+\tilde{V}_c\right) + \left(\frac{V'_c}{\tan^2 \tilde{\theta}}+\tilde{V'}_c\right) \delta U + \left(\frac{V''_c}{\tan^2 \tilde{\theta}}+\tilde{V''}_c\right) \frac{(\delta U)^2}{2}.
\end{equation}
Following the same argument presented in the previous section, we must have $\tilde{C}_1 >0$. If this is the case, then imposing that \eqref{eq:lagrangeanimang} has an imaginary part, summing all such contributions of $\mathcal{L}_j$ in \eqref{eq:thermalflucpartang2} and finally taking the continuum limit we arrive at a new and explicit analytical expression for $\mathrm{Im} V_{Q\bar{Q}}$ valid for a large class of gravity duals
\begin{equation}
\label{eq:ImFQQang}
\mathrm{Im}\,V_{Q\bar{Q}} = -\frac{1}{4\alpha'}\frac{1}{\sqrt{\tilde{C}_1}} \left[ \frac{\left(\frac{V_c'}{\tan^2\tilde{\theta}}+ \tilde{V}_c'\right)^2}{2 \left(\frac{V_c''}{\tan^2\tilde{\theta}}+ \tilde{V}_c''\right)} - \left(\frac{V_c}{\tan^2\tilde{\theta}}+ \tilde{V}_c\right) \right]\,.
\end{equation}
For this expression to be valid, we must impose $\mathrm{Im} V_{Q\bar{Q}}<0$. Also, note that to compute $\tilde{C}_1$ using \eqref{eq:C1tilde} we need to use two pieces of information concerning the shape of $\bar{U}(\sigma)$ at $\sigma \sim 0$, $\bar{U}''(0)$ and $b$.

\subsection{An explicit example - Thermal $\mathcal{N}=4$ SYM}

Let us apply the results of the foregoing sections to strongly coupled thermal $\mathcal{N}=4$ SYM. The metric is given by equation \eqref{eq:metricads}, and the $M,N,P,V$ functions are the same as in \eqref{eq:MVPNfuncAdS}. Evaluation of the equations of motion \eqref{eq:eqmotionAfin} and \eqref{eq:eqmotionBfin} lead to
\begin{equation}
\label{eq:eqmovangAdSA}
q^2 \left(\frac{dy}{d\tilde{\sigma}}\right)^2 = (y^4 - \cosh^2\eta)(y^4-1-p^2) - q^2(y^4-1) \quad \quad \mathrm{and}
\end{equation}
\begin{equation}
\label{eq:eqmovangAdSB}
\left(\frac{dz}{d\tilde{\sigma}}\right)^2 = \frac{p^2}{q^2} \left(\frac{y^4 - \cosh^2\eta}{y^4-1}^2 \right)^2,
\end{equation}
where we defined the dimensionless variables $y\equiv U/U_h$, $z\equiv X_d U_h/R^2$ and $\tilde{\sigma} \equiv \sigma U_h/R^2$ as well as the dimensionless integration constants $q^2 \equiv R^4 Q^2/U_h^4$ and $p^2 = R^4 K^2/U_h^4$. With these variables, the boundary conditions \eqref{eq:boundangle} become
\begin{align}
\label{eq:boundangle}
y \left(\pm \pi \frac{LT}{2} \sin \theta \right) & = \tilde{\Lambda} \nonumber \\
z \left(\pm \pi \frac{LT}{2} \sin \theta \right) & = \pm \pi \frac{LT}{2} \cos \theta
\end{align}
where we substituted $L$ for the dimensionless variable $LT = L U_h/(\pi R^2)$ to explicit show the conformal character of the gauge theory and $\tilde{\Lambda}$ is the dimensionless UV cutoff. Using the equations above we arrive at expressions relating the integration constants $q,p$ with the physical parameters $LT, \theta$:
\begin{equation}
\label{eq:LangAdS1}
\frac{LT}{2} \pi \sin \theta  =  q \int_{y_c}^{\tilde{\Lambda}} \, \frac{dy}{\sqrt{(y^4-1-p^2)(y^4-\cosh^2\eta)-q^2(y^4-1)}} \quad \quad \mathrm{and}
\end{equation}
\begin{equation}
\label{eq:LangAdS2}
\frac{LT}{2} \pi \cos \theta  =  p \int_{y_c}^{\tilde{\Lambda}} \, dy \, \frac{y^4-\cosh^2\eta}{y^4-1} \frac{1}{\sqrt{(y^4-1-p^2)(y^4-\cosh^2\eta)-q^2(y^4-1)}}\,.
\end{equation}
The condition \eqref{eq:relationUc} reduces to
\begin{equation}
\label{eq:conditionangAdS}
(y_c^4-1-p^2)(y_c^4-\cosh^2\eta)-q^2(y_c^4-1) = 0\,.
\end{equation}
The real part of the heavy quark potential, after regularization, is given by \eqref{eq:actionangreg}. This yields
\begin{equation}
\label{eq:ReFQQangregAdS}
\frac{\mathrm{Re} \, V_{Q\bar{Q}}}{T} = \sqrt{\lambda} \int_{y_c}^{\infty} dy \left[ \frac{y^4-\cosh^2\eta}{\sqrt{(y^4-\cosh^2\eta)(y^4-1-p^2)-q^2(y^4-1)}}-1 \right] - \sqrt{\lambda} (y_c -1)\,.
\end{equation}
These results agree with those in \cite{Liu:2006he}. As for the imaginary part, we use \eqref{eq:ImFQQang}
\begin{equation}
\label{eq:ImFQQangregAdS}
\frac{\mathrm{Im} \, V_{Q\bar{Q}}}{T} = -\frac{\pi \sqrt{\lambda}}{12} \frac{[3( \cos^2 \tilde{\theta} + \cosh^2 \eta \sin^2 \tilde{\theta} ) - y_c^4]}{\sqrt{y''(0)^2 \left( \frac{y_c^4-\cosh^2 \eta}{y_c^4-1} \right) + \frac{2 y_c^3 y''(0)}{\sin \tilde{\theta}} + \frac{6 \tilde{b}}{\tan \tilde{\theta}} (y_c^4-1)}},
\end{equation}
where we defined $\tilde{b} = U_h^2 b/R^4$. For this formula to be valid, we must guarantee that $\mathrm{Im} \, V_{Q\bar{Q}} <0$. In terms of $\tilde{\sigma}$, the expansion \eqref{eq:XdexpansionAdS} takes the form
\begin{equation}
\label{eq:XdexpansionAdS}
z(\tilde{\sigma}) = \frac{\tilde{\sigma}}{\tan \tilde{\theta}} + \tilde{b} \tilde{\sigma}^3 + O (\tilde{\sigma}^5)\,.
\end{equation}

With these expressions, we can proceed to solve the equations numerically. We consider $\eta$ and $\theta$ as given and fixed. The first step is to relate $(q,p)$ with $(LT,\theta)$. First, we solve \eqref{eq:conditionangAdS} for $y_c$ as a function of $q$ and $p$
\begin{align}
\label{eq:ycqp}
y_c (q,p) = & \left[ \frac{1}{2} \sqrt{\left(-\cosh ^2(\eta )-p^2-q^2-1\right)^2-4
   \left(p^2 \cosh ^2(\eta )+\cosh ^2(\eta )+q^2\right)} + \right. \nonumber \\ & \left. + \frac{\cosh
   ^2(\eta )}{2}+\frac{p^2}{2}+\frac{q^2}{2}+\frac{1}{2} \right]^{1/4}\,.
\end{align}
Then, we substitute \eqref{eq:ycqp} into equations \eqref{eq:LangAdS1} and \eqref{eq:LangAdS2}. This results in two equations relating $(q,p)$ to $(L,\theta)$. However, the ratio of these equations involves only $(q,p)$ and $\theta$. Thus, for fixed $\theta$, we can solve this equation numerically for $p$ as a function of $q$, $p(q)$. Therefore, $y_c$ is now a function only of $p$, $y_c = y_c(q,p(q))$. We can insert $p(q)$ and $y_c(q,p(q))$ into any of the equations \eqref{eq:LangAdS1} or \eqref{eq:LangAdS2} to solve for $LT$ as a function of $q$. Finally, we can obtain $\mathrm{Re} \, V_{Q\bar{Q}}$ as a function of $LT$ by considering the parametric curve $(LT(q),\mathrm{Re} \, V_{Q\bar{Q}}(q))$.

\begin{figure}[h!tp]
\begin{center}

  \begin{subfigure}[b]{0.45\textwidth}
  \label{fig:pqvareta}
  \includegraphics[width=\textwidth]{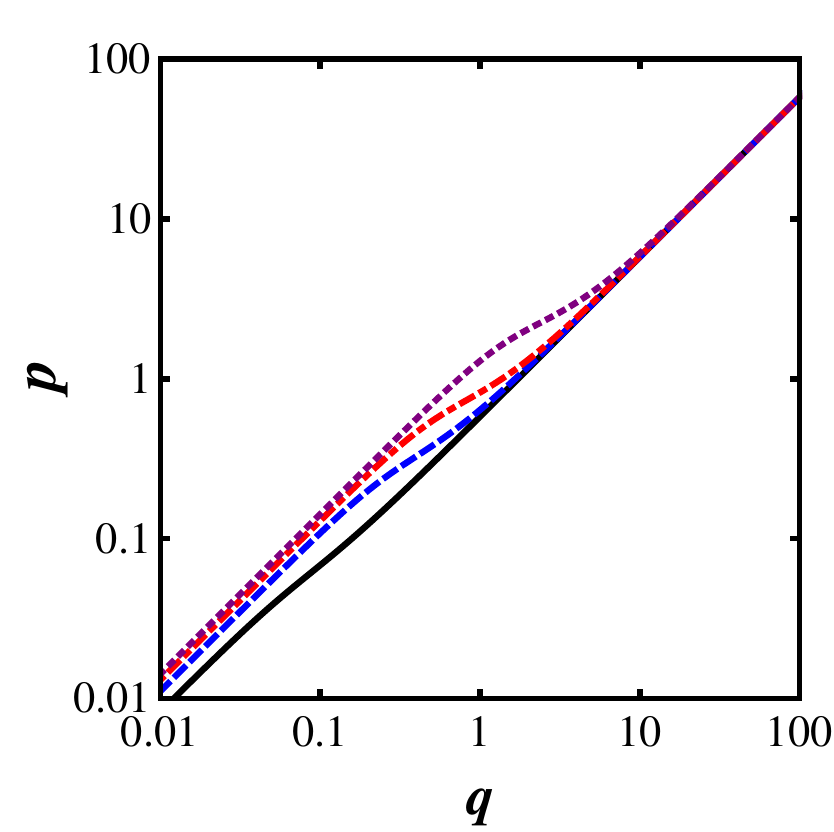}
  \end{subfigure}
  \begin{subfigure}[b]{0.47\textwidth}
  \label{fig:pqvartheta}
  \includegraphics[width=\textwidth]{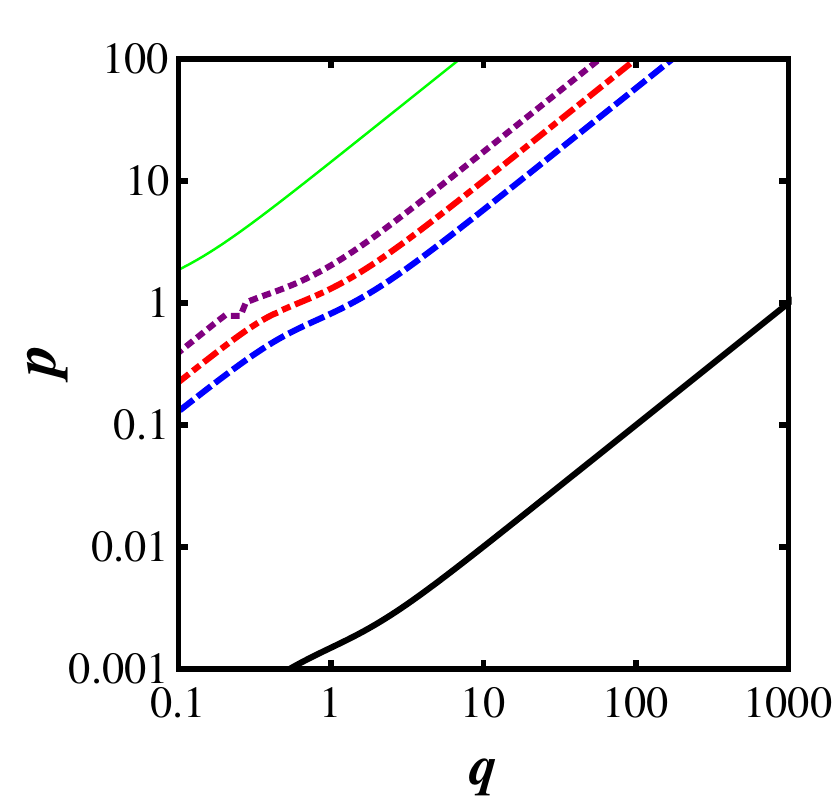}
  \end{subfigure}
  
  \caption{(Color online) The constant $p$ as a function of $q$, (left plot) for $\theta = \pi/3$ and $\eta = 0.1$ (thick black curve), $\eta = 0.5$ (dashed blue curve), $\eta = 1.0$ (dashed-dotted red curve) and $\eta = 2.0$ (dotted purple curve) and (right plot) for $\eta = 1.0$ and $\theta = \pi/2$ (thick black curve), $\theta = \pi/3$ (dashed blue curve), $\theta = \pi/4$ (dashed-dotted red curve), $\theta = \pi/6$ (dotted purple curve) and $\theta = \pi/45$ (thin green curve). The calculations are done for the case of a strongly coupled $\mathcal{N}=4$ SYM plasma.}
  \label{fig:pqang}

\end{center}
\end{figure}

In Fig.\ \ref{fig:pqang} we present the numerical results for $p$ as a function of $q$, for a fixed angle and some choices of the rapidity $\eta$ and for fixed $\eta$ and some choices of $\theta$, reproducing (and extending) the results in \cite{Liu:2006nn,Liu:2006he}. We see that $p$ is a monotonically increasing function of $q$, as it should be for the problem to be well-posed. 

\begin{figure}[h!tp]
\begin{center}
  \begin{subfigure}[b]{0.45\textwidth}
  \label{fig:LTqvareta}
  \includegraphics[width=\textwidth]{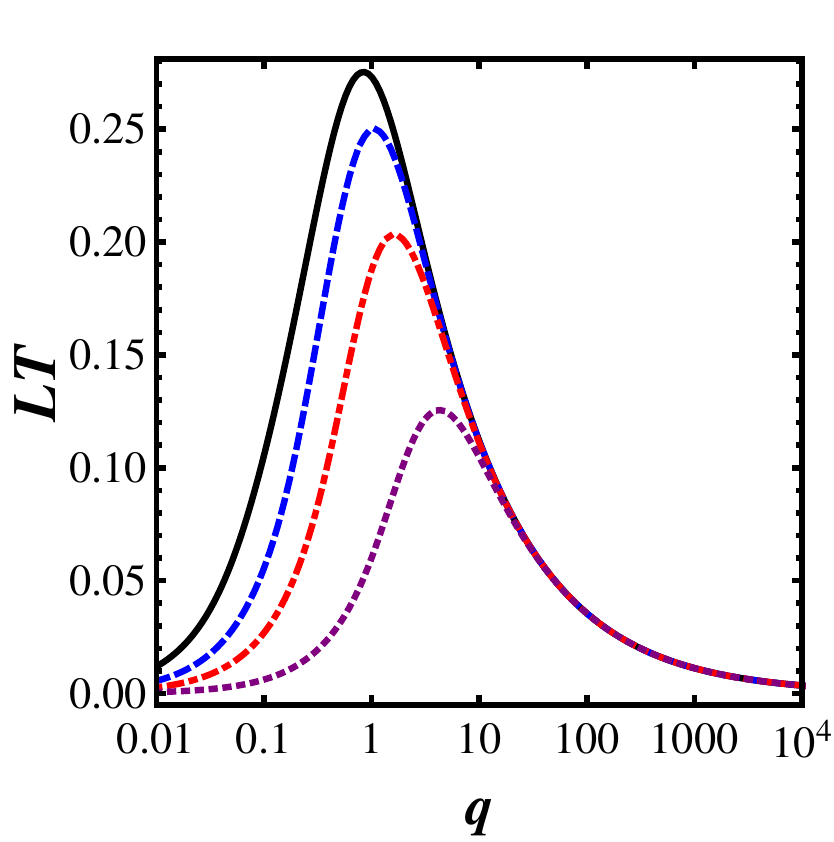}
  \end{subfigure}
  \begin{subfigure}[b]{0.45\textwidth}
  \label{fig:LTqvartheta}
  \includegraphics[width=\textwidth]{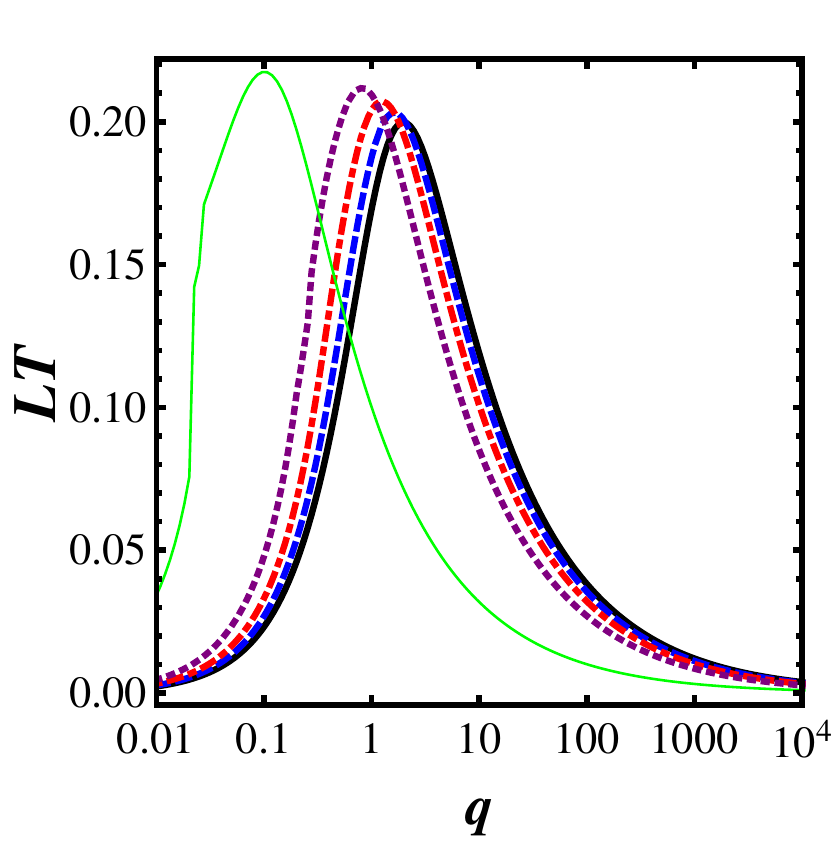}
  \end{subfigure}
  
  \caption{(Color online) $LT$ as a function of $q$, (left plot) for $\theta = \pi/3$ and $\eta = 0.1$ (thick black curve), $\eta = 0.5$ (dashed blue curve), $\eta = 1.0$ (dash-dotted red curve) and $\eta = 2.0$ (dotted purple curve); and (right plot) for $\eta = 1.0$ and $\theta = \pi/2$ (thick black curve), $\theta = \pi/3$ (dashed blue curve), $\theta = \pi/4$ (dashed-dotted red curve), $\theta = \pi/6$ (dotted purple curve) and $\theta = \pi/45$ (thin green curve). The calculations are done for the case of a strongly coupled $\mathcal{N}=4$ SYM plasma.}
  \label{fig:LTang}
\end{center}
\end{figure}

In Fig.\ \ref{fig:LTang} we show $LT$ as a function of $q$. We see that $LT$ assumes a maximum value, $LT_{max}$ which depends strongly on the rapidity $\eta$, for a fixed orientation of the dipole, but only slightly on the angle $\theta$, with $\eta$ fixed. As discussed in the perpendicular case, we will not take $LT_{max}$ to define the dissociation length, but only as an indicative of the limit of validity of our classical gravity calculation. 

\begin{figure}[h!tp]
\begin{center}

  \begin{subfigure}[b]{0.45\textwidth}
  \label{fig:ReFQQvareta}
  \includegraphics[width=\textwidth]{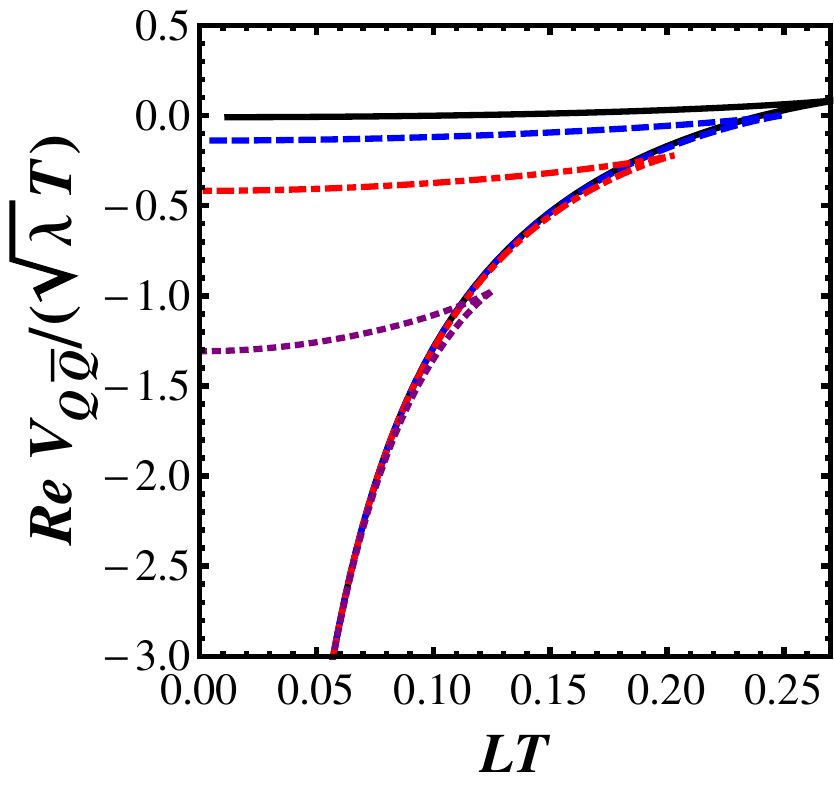}
  \end{subfigure}
  \begin{subfigure}[b]{0.47\textwidth}
  \label{fig:ReFQQvartheta}
  \includegraphics[width=\textwidth]{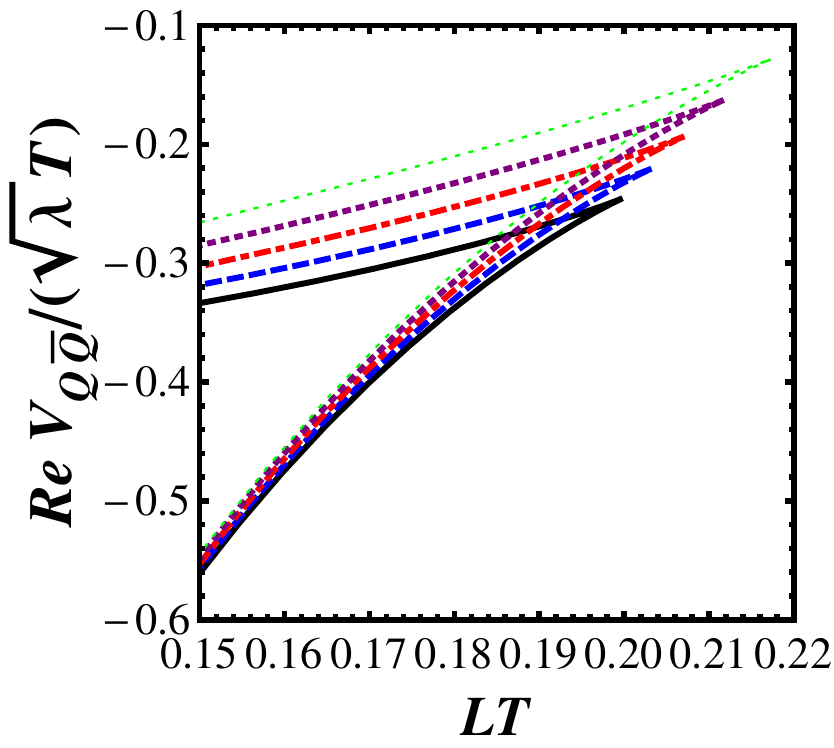}
  \end{subfigure}
  
  \caption{(Color online) $\mathrm{Re} \, V_{Q\bar{Q}}$ as a function of $q$, (left plot) for $\theta = \pi/3$ and $\eta = 0.1$ (thick black curve), $\eta = 0.5$ (dashed blue curve), $\eta = 1.0$ (dashed-dotted red curve) and $\eta = 2.0$ (dotted purple curve); and for $\eta = 1.0$ and $\theta = \pi/2$ (thick black curve), $\theta = \pi/3$ (dashed blue curve), $\theta = \pi/4$ (dashed-dotted red curve), $\theta = \pi/6$ (dotted purple curve) and $\theta = \pi/45$ (thin green curve). In (right plot) we show only the detail of $\mathrm{Re} \, V_{Q\bar{Q}}$ around $LT_{max}$. The calculations are done for the case of a strongly coupled $\mathcal{N}=4$ SYM plasma.}
  \label{fig:ReFQQang}
\end{center}
\end{figure}

In Fig.\ \ref{fig:ReFQQang} we present $\mathrm{Re} \, V_{Q\bar{Q}}$ as a function of $LT$. The region of small $LT$ does not change appreciably with either the rapidity $\eta$ or the angle $\theta$, as it should be expected, since for small temperatures or short distances the interaction of the $Q\bar{Q}$ pair with the plasma should not be relevant. Near $LT_{max}$, $\mathrm{Re} \, V_{Q\bar{Q}}$ varies slightly with $\eta$ and almost nothing at all with $\theta$ (note that in Fig.\ \ref{fig:ReFQQang} (right plot) we are considering only a small region of $LT$ to zoom in the effect of varying $\theta$). The upper, unphysical, branch in the Fig.\ \ref{fig:ReFQQang} represents the region $0 < q < q_{max}$, where $q_{max}$ is the value of $q$ which gives $LT_{max}$: $LT_{max} = q_{max}$. The lower branch, which is the dominant contribution to the action, is given by the region with $q>q_{max}$.

\begin{figure}[h!tp]
\begin{center}
  \begin{subfigure}[b]{0.45\textwidth}
  \label{fig:profiley}
  \includegraphics[width=\textwidth]{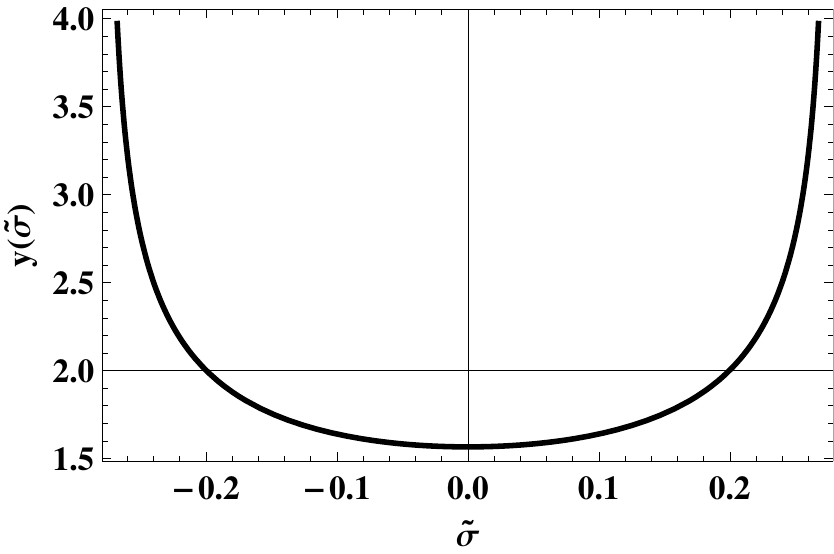}
  \end{subfigure}
  \begin{subfigure}[b]{0.47\textwidth}
  \label{fig:profilez}
  \includegraphics[width=\textwidth]{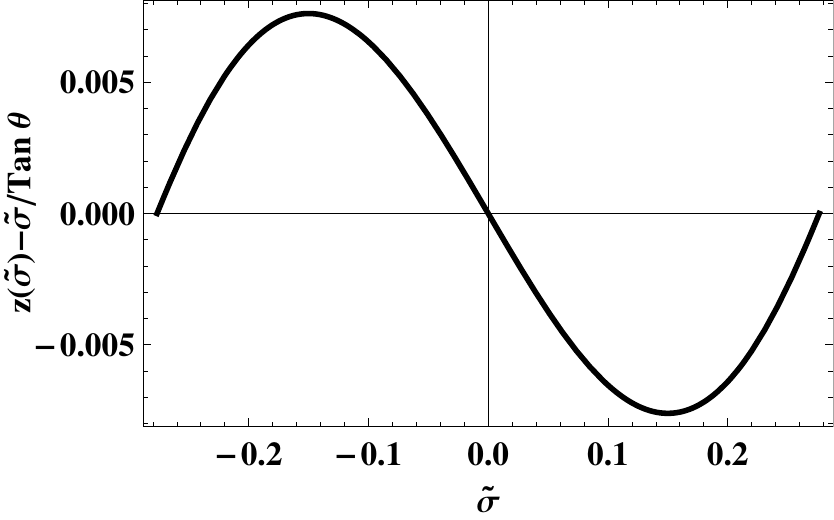}
  \end{subfigure}
  
  \caption{The profile of the string for an $\mathcal{N}=4$ SYM plasma. (left plot) $y$ as a function of $\tilde{\sigma}$. (right plot) $z- \tilde{\sigma}/\tan \theta$ as a function of $\tilde{\sigma}$ - the subtraction of $\tilde{\sigma}/\tan \theta$ is to remove the trivial shape the string would assume if its projection on the $(X_1,X_{d-1})$ plane were a straight line segment joining the endpoints of the dipole. These figures refer to a situation with $\theta = \pi/3$, $\eta = 1.0$, and $LT = 0.203$. }
  \label{fig:profile}
\end{center}
\end{figure}

With the real part under control, let us move on to compute $\mathrm{Im} \, V_{Q\bar{Q}}$ for this case. Before using \eqref{eq:ImFQQangregAdS} to calculate $\mathrm{Im} \, V_{Q\bar{Q}}$, we must obtain $\tilde{\theta}$, $y''(0)$ and $\tilde{b}$ for the $LT$ in consideration, for fixed $\theta$ and $\eta$. The effective angle $\tilde{\theta}$ is obtained directly from \eqref{eq:eqmovangAdSB} evaluated at $\tilde{\sigma} = 0$, $y = y_c$ (noting that $z'(0) = 1/\tilde{\theta}$, from \eqref{eq:XdexpansionAdS}). To evaluate $y''(0)$ and $\tilde{b}$ we solve the equations of motion \eqref{eq:eqmovangAdSA} and \eqref{eq:eqmovangAdSB} subject to the boundary conditions $y(0) = y_c$ and $z(0) = 0 $. In principle we could use the set \eqref{eq:boundangle} of boundary conditions, but since these conditions refer to the string far from $\tilde{\sigma} = 0$, exactly the region of interest for the evaluation of $y''(0) = 0$ and $z'''(0) = 0$, they are not very useful. An example of the shape of the string when we solve these equations numerically is given in Fig.\ \eqref{fig:profile} - we note in this figure we subtracted the contribution that would appear if the projection of the string on $(X_1,X_D)$ were a straight line joining the endpoints of the $\bar{Q}Q$ dipole. After figuring out the shape of the string, we may evaluate $y''(0)$ and $\tilde{b}$.

\begin{figure}[h!tp]
\begin{center}
\label{fig:ImFQQvareta}
  \begin{subfigure}[b]{0.45\textwidth}
  \includegraphics[width=\textwidth]{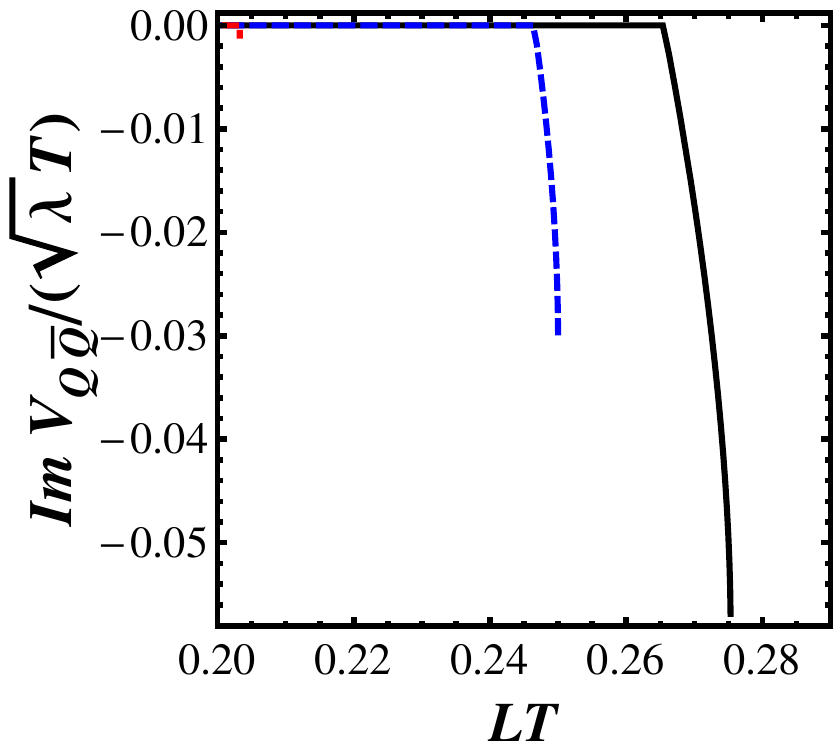}
  \end{subfigure}
  \begin{subfigure}[b]{0.48\textwidth}
  \label{fig:ImFQQvartheta}
  \includegraphics[width=\textwidth]{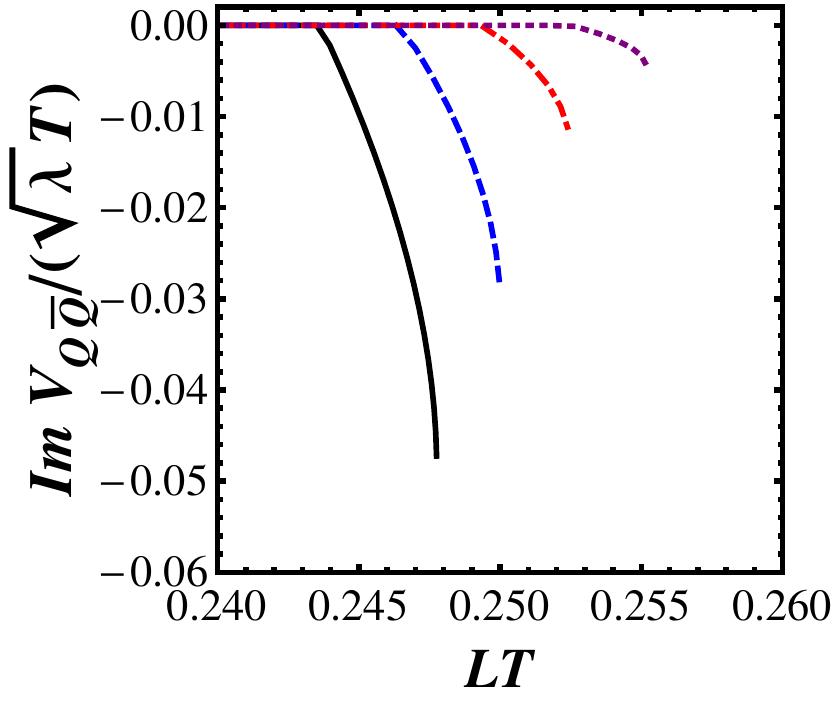}
  \end{subfigure}
  
  \caption{(Color online) $\mathrm{Im} \, V_{Q\bar{Q}}/(\sqrt{\lambda}T)$ as a function of $q$, (left panel) for $\theta = \pi/3$ and $\eta = 0.1$ (solid black curve), $\eta = 0.5$ (dashed blue curve) and $\eta =1.0$ (dotted-dashed red curve) and (right panel) for $\eta = 0.5$ and $\theta = \pi/2$ (solid black curve), $\theta = \pi/3$ (dashed blue curve), $\theta = \pi/4$ (dotted-dashed red curve) and $\theta = \pi/6$ (dotted purple curve).}
  \label{fig:ImFQQang}
\end{center}
\end{figure}

\begin{figure}[h!t]
\begin{center}
\includegraphics[width=0.6 \textwidth]{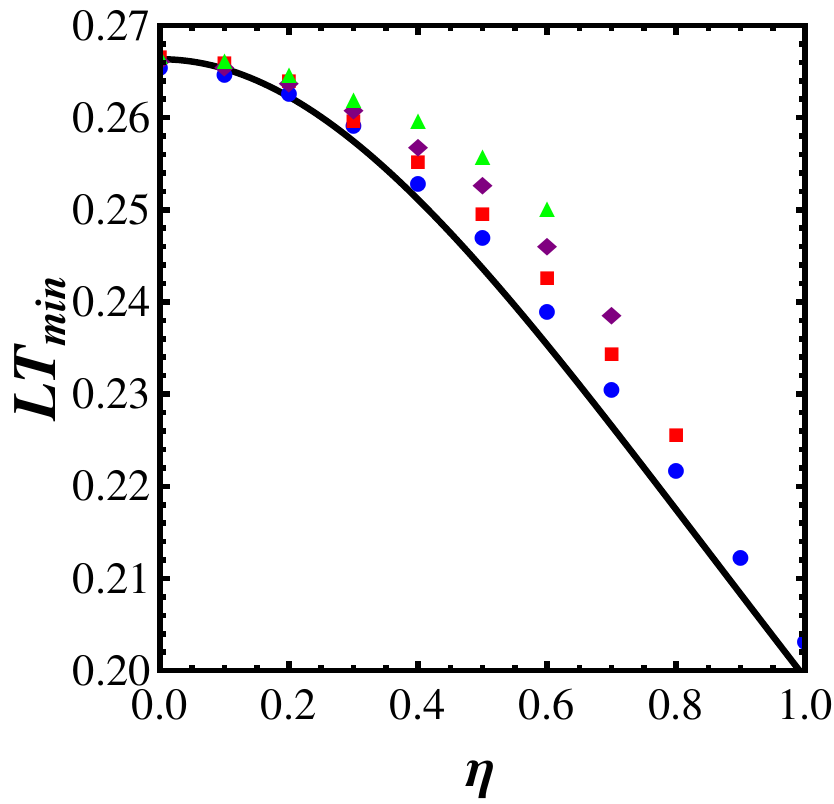}
\caption{$LT_{min}$ for various angles $\theta$, as a function of the rapidity $\eta$, for a $Q\bar{Q}$ pair moving through an $\mathcal{N}=4$ SYM plasma. The solid black curve corresponds to $\theta = \pi/2$, $\theta = \pi/3$ (blue circles), $\theta = \pi/4$ (red squares), $\theta = \pi/6$ (purple diamonds), and $\theta = \pi/45$ (green triangles). }
\label{fig:LTminvartheta}
\end{center}
\end{figure}

With $y''(0)$ and $\tilde{b}$ known, considering only $LT<LT_{max}$ and ensuring that \eqref{eq:ImFQQangregAdS} is negative, we can calculate $\mathrm{Im} \, V_{Q\bar{Q}}$ as a function of $q$. As we know $LT(q)$, we can plot $\mathrm{Im} \, V_{Q\bar{Q}}/T$ as a function of $LT$. In Fig.\ \ref{fig:ImFQQang} we show the results of these calculations. We see that, for a fixed $\theta \neq \pi/2$, increasing $\eta$ decreases the interval of $LT$ allowed for the calculation. Such a behavior is also confirmed when we investigate the $LT_{min}$ for the onset of the imaginary part, as shown in Fig.\ \ref{fig:LTminvartheta}. We see that $LT_{min}$ decreases strongly with $\eta$; there is also a slight increase for decreasing $\theta$. When we fix $\eta$ and vary $\theta$ (Fig.\ \ref{fig:ImFQQang} - right panel), we see that the region where the calculation is valid decreases and $\mathrm{Im} \, V_{Q\bar{Q}}/T$ is smaller as $\theta$ decreases from the perpendicular case $\theta = \pi/2$ to $\theta \to 0$. This suggests that a dipole oriented parallel to the wind should have a smaller thermal width and the interactions between the heavy quark-antiquark pair are less screened by the plasma in comparison to the perpendicular case. 

\section{Conclusions}
\label{sec:concl}

In this work, we have used the thermal worldsheet fluctuation method \cite{Noronha:2009da,Finazzo:2013rqy} to investigate the imaginary part $\mathrm{Im} \, V_{Q\bar{Q}}$ of moving heavy quarkonia in strongly coupled plasmas with gravity duals. We have developed a general formalism to holographically compute $\mathrm{Im} \, V_{Q\bar{Q}}$ in the case where the heavy quark dipole has an arbitrary orientation with respect to the plasma velocity of the underlying strongly coupled plasma. The general formula for this quantity is shown in \eqref{eq:ImFQQang}. Also, we have discussed in detail the regime of validity of the method in this case with nonzero rapidity, greatly expanding the original discussion in \cite{Finazzo:2013rqy}. In fact, we found that the requirement for small thermal fluctuations around the classical worldsheet solution imposes severe constraints in the calculation of $\mathrm{Im} \, V_{Q\bar{Q}}$. The most important constraint in this case is that only slowly moving quarkonia can be considered. In fact, for large rapidities the range of applicability of the saddle point approximation used in the calculation of the imaginary part of the potential decreases significantly. Therefore, the case corresponding to moderately large quarkonia rapidities would require to go beyond the saddle point approximation in the calculation of the string path integral for the worldsheet fluctuations. This challenging task is beyond the scope of the present study and we hope to address this problem in a future work. It would be interesting to see if the other holographic methods to determine $\mathrm{Im} \, V_{Q\bar{Q}}$ \cite{Albacete:2009bp, Hayata:2012rw} are better suited to tackle the problem of rapidly moving quarkonia (this is the case of the method used in \cite{Faulkner:2008qk}).

We have applied the worldsheet fluctuation method to evaluate $\mathrm{Im} \, V_{Q\bar{Q}}$ for heavy quarkonia moving through a strongly coupled $\mathcal{N}=4$ SYM plasma. When the velocity becomes parallel ($\theta \to 0$) to the heavy quark dipole, we find that the imaginary part of the potential becomes smaller than that in the perpendicular case ($\theta=\pi/2$). This shows that at strong coupling the anisotropy induced by the nonzero rapidity of the heavy quark pair becomes a relevant factor in the calculation of the associated thermal width in the plasma, which is suppressed for small angles.

We found a strong dependence of $\mathrm{Im} \, V_{Q\bar{Q}}$ on the rapidity and, taking into account the strict consistency constraints imposed in our holographic setup, for increasing $\eta$ the onset of the imaginary part of the potential occurs at smaller values of $LT$, though the precise magnitude of this quantity cannot be reliably determined at the moment with our approximations. Our results indicate that moving quarkonia are indeed less stable in a strongly coupled plasma, which is consistent with previous findings using other approaches \cite{Liu:2006nn,Liu:2006he,Ejaz:2007hg,Faulkner:2008qk}.

It would be interesting to extend the calculations performed here to other gravity backgrounds, such as those that display a confinement-deconfinement transition described by bottom-up Einstein + Scalar models \cite{Gursoy:2007cb,Gursoy:2007er,Gursoy:2008bu,Gubser:2008ny,Noronha:2009ud}. 

Note added: When this article was being finished, we became aware of Ref.\ \cite{Ali-Akbari:2014vpa} which also discussed the effects of nonzero rapidity on the imaginary part of the heavy quark potential. In their work, they computed $\mathrm{Im} \, V_{Q\bar{Q}}$ only for parallel and perpendicular configurations. In our work we not only considered arbitrary orientations of the dipole with respect to the hot plasma wind but also discussed in detail the regime of applicability of the calculations done using the saddle point approximation (this critical analysis was not performed in \cite{Ali-Akbari:2014vpa}). 

\acknowledgements

This work was supported by Funda\c c\~ao de Amparo \`a Pesquisa do Estado de
S\~ao Paulo (FAPESP) and Conselho Nacional de Desenvolvimento Cient\'ifico e
Tecnol\'ogico (CNPq).



\begin{thebibliography}{99}

\bibitem{Shuryak:1980tp} 
  E.~V.~Shuryak,
  Phys.\ Rept.\  {\bf 61}, 71 (1980).

\bibitem{Karsch:1987pv} 
  F.~Karsch, M.~T.~Mehr and H.~Satz,
  Z.\ Phys.\ C {\bf 37}, 617 (1988).

\bibitem{Wilson:1974sk} 
  K.~G.~Wilson,
  Phys.\ Rev.\ D {\bf 10}, 2445 (1974).

 \bibitem{wilsonloop}
 J.~-L.~Gervais and A.~Neveu, Nucl.\ Phys.\ B {\bf 163}, 189 (1980); A.~M.~Polyakov, Nucl.\ Phys.\ B {\bf 164}, 171 (1980).

  
\bibitem{imvrefs} M.~Laine et al., JHEP {\bf 0703} (2007) 054; JHEP {\bf 0705}, 028
(2007); A.~Beraudo, J.~P.~Blaizot and C.~Ratti, Nucl.\ Phys.\ A {\bf 806}, 312 (2008); N.~Brambilla, J.~Ghiglieri, A.~Vairo, P.~Petreczky, Phys.
Rev. D {\bf 78}, 014017 (2008).

\bibitem{otherrefs1} C.~Miao, A.~Mocsy, and P.~Petreczky, Nucl.\ Phys.\
A {\bf 855}, 125 (2011), arXiv:1012.4433 [hep-ph]; N.~Brambilla, M.~A.~Escobedo, J.~Ghiglieri, J.~Soto, and A.~Vairo,
JHEP {\bf 09}, 038 (2010), arXiv:1007.4156 [hep-ph].

\bibitem{otherrefsImV} Y.~Burnier, M.~Laine, and M.~Vepsalainen, (2009),
arXiv:0903.3467 [hep-ph]; A.~Dumitru, Y.~Guo,
and M.~Strickland, Phys.\ Rev.\ D {\bf 79}, 114003 (2009),
arXiv:0903.4703 [hep-ph]; O.~Philipsen and M.~Tassler,
(2009), arXiv:0908.1746 [hep-ph].

\bibitem{Laine:2006ns} 
  M.~Laine, O.~Philipsen, P.~Romatschke and M.~Tassler,
  JHEP {\bf 0703}, 054 (2007)
  [hep-ph/0611300].
  
\bibitem{Rothkopf:2011db} 
  A.~Rothkopf, T.~Hatsuda and S.~Sasaki,
  Phys.\ Rev.\ Lett.\  {\bf 108}, 162001 (2012)
  [arXiv:1108.1579 [hep-lat]].
  
\bibitem{Aarts:2011sm} 
  G.~Aarts, C.~Allton, S.~Kim, M.~P.~Lombardo, M.~B.~Oktay, S.~M.~Ryan, D.~K.~Sinclair and J.~I.~Skullerud,
  JHEP {\bf 1111}, 103 (2011)
  [arXiv:1109.4496 [hep-lat]].
  
\bibitem{Aarts:2013kaa} 
  G.~Aarts, C.~Allton, S.~Kim, M.~P.~Lombardo, S.~M.~Ryan and J.~-I.~Skullerud,
  JHEP {\bf 1312}, 064 (2013)
  [arXiv:1310.5467 [hep-lat]].
  
  


  
\bibitem{Noronha:2009da} 
  J.~Noronha and A.~Dumitru,
  Phys.\ Rev.\ Lett.\  {\bf 103}, 152304 (2009)
  [arXiv:0907.3062 [hep-ph]].

\bibitem{Albacete:2009bp} 
  J.~L.~Albacete, Y.~V.~Kovchegov and A.~Taliotis,
  Phys.\ Rev.\ D {\bf 78}, 115007 (2008)
  [arXiv:0807.4747 [hep-th]].
  
  
\bibitem{Hayata:2012rw} 
  T.~Hayata, K.~Nawa and T.~Hatsuda,
  arXiv:1211.4942 [hep-ph].

\bibitem{Finazzo:2013rqy} 
  S.~I.~Finazzo and J.~Noronha,
  JHEP {\bf 1311}, 042 (2013)
  [arXiv:1306.2613 [hep-ph]].

\bibitem{Fadafan:2013bva} 
  K.~B.~Fadafan, D.~Giataganas and H.~Soltanpanahi,
  arXiv:1306.2929 [hep-th].
  
\bibitem{Fadafan:2013coa} 
  K.~B.~Fadafan and S.~K.~Tabatabaei,
  Eur.\ Phys.\ J.\ C {\bf 74}, 2842 (2014)
  [arXiv:1308.3971 [hep-th]].
  
\bibitem{Akamatsu:2011se} 
  Y.~Akamatsu and A.~Rothkopf,
  Phys.\ Rev.\ D {\bf 85}, 105011 (2012)
  [arXiv:1110.1203 [hep-ph]].
  
\bibitem{Akamatsu:2014qsa} 
  Y.~Akamatsu,
  arXiv:1403.5783 [hep-ph].
  
  
\bibitem{Strickland:2011mw} 
  M.~Strickland,
  Phys.\ Rev.\ Lett.\  {\bf 107}, 132301 (2011)
  [arXiv:1106.2571 [hep-ph]].
  
\bibitem{Strickland:2011aa} 
  M.~Strickland and D.~Bazow,
  Nucl.\ Phys.\ A {\bf 879}, 25 (2012)
  [arXiv:1112.2761 [nucl-th]].
  
\bibitem{mikenew}
M.~Margotta, K.~McCarty, C.~McGahan, M.~Strickland,
and D.~Yager-Elorriaga, Phys.\ Rev.\ D {\bf 83}, 105019 (2011),
arXiv:1101.4651 [hep-ph].

\bibitem{Aarts:2012ka} 
  G.~Aarts, C.~Allton, S.~Kim, M.~P.~Lombardo, M.~B.~Oktay, S.~M.~Ryan, D.~K.~Sinclair and J.~-I.~Skullerud,
  JHEP {\bf 1303}, 084 (2013)
  [arXiv:1210.2903 [hep-lat]].

\bibitem{Escobedo:2013tca} 
  M.~A.~Escobedo, F.~Giannuzzi, M.~Mannarelli and J.~Soto,
  Phys.\ Rev.\ D {\bf 87}, no. 11, 114005 (2013)
  [arXiv:1304.4087 [hep-ph]].
  
  

\bibitem{Maldacena:1997re} 
  J.~M.~Maldacena,
  Adv.\ Theor.\ Math.\ Phys.\  {\bf 2}, 231 (1998)
  [Int.\ J.\ Theor.\ Phys.\  {\bf 38}, 1113 (1999)]
  [hep-th/9711200]
  
\bibitem{Witten:1998qj} 
  E.~Witten,
  Adv.\ Theor.\ Math.\ Phys.\  {\bf 2}, 253 (1998)
  [hep-th/9802150].
  
\bibitem{Witten:1998zw} 
  E.~Witten,
  Adv.\ Theor.\ Math.\ Phys.\  {\bf 2}, 505 (1998)
  [hep-th/9803131].

\bibitem{Maldacena:1998im} 
  J.~M.~Maldacena,
  Phys.\ Rev.\ Lett.\  {\bf 80}, 4859 (1998)
  [hep-th/9803002].
  
\bibitem{Brandhuber:1998bs} 
  A.~Brandhuber, N.~Itzhaki, J.~Sonnenschein and S.~Yankielowicz,
  Phys.\ Lett.\ B {\bf 434}, 36 (1998)
  [hep-th/9803137].
 
\bibitem{Rey:1998bq} 
  S.~-J.~Rey, S.~Theisen and J.~-T.~Yee,
  Nucl.\ Phys.\ B {\bf 527}, 171 (1998)
  [hep-th/9803135].

\bibitem{Kinar:1998vq} 
  Y.~Kinar, E.~Schreiber and J.~Sonnenschein,
  Nucl.\ Phys.\ B {\bf 566}, 103 (2000)
  [hep-th/9811192].

\bibitem{Sonnenschein:1999if} 
  J.~Sonnenschein,
  hep-th/0003032.
  
\bibitem{Liu:2006nn} 
  H.~Liu, K.~Rajagopal and U.~A.~Wiedemann,
  Phys.\ Rev.\ Lett.\  {\bf 98}, 182301 (2007)
  [hep-ph/0607062].
  
\bibitem{Liu:2006he} 
  H.~Liu, K.~Rajagopal and U.~A.~Wiedemann,
  JHEP {\bf 0703}, 066 (2007)
  [hep-ph/0612168].
  
\bibitem{Caceres:2006ta} 
  E.~Caceres, M.~Natsuume and T.~Okamura,
  JHEP {\bf 0610}, 011 (2006)
  [hep-th/0607233].
  
\bibitem{Faulkner:2008qk} 
  T.~Faulkner and H.~Liu,
  Phys.\ Lett.\ B {\bf 673}, 161 (2009)
  [arXiv:0807.0063 [hep-th]].
  
\bibitem{Ejaz:2007hg} 
  Q.~J.~Ejaz, T.~Faulkner, H.~Liu, K.~Rajagopal and U.~A.~Wiedemann,
  JHEP {\bf 0804}, 089 (2008)
  [arXiv:0712.0590 [hep-th]].
  
\bibitem{gradshteyn}
S. Gradshteyn and I.M. Ryzhik; A. Jeffrey, D. Zwillinger, editors. \emph{Table of Integrals, Series, and Products}, seventh edition. Academic Press, 2007.

\bibitem{Karch:2002sh} 
  A.~Karch and E.~Katz,
  JHEP {\bf 0206}, 043 (2002)
  [hep-th/0205236].

\bibitem{Bak:2007fk} 
  D.~Bak, A.~Karch and L.~G.~Yaffe,
  JHEP {\bf 0708}, 049 (2007)
  [arXiv:0705.0994 [hep-th]].

\bibitem{Gursoy:2007cb} 
  U.~Gursoy and E.~Kiritsis,
  JHEP {\bf 0802}, 032 (2008)
  [arXiv:0707.1324 [hep-th]].
  
\bibitem{Gursoy:2007er} 
  U.~Gursoy, E.~Kiritsis and F.~Nitti,
  JHEP {\bf 0802}, 019 (2008)
  [arXiv:0707.1349 [hep-th]].

\bibitem{Gursoy:2008bu} 
  U.~Gursoy, E.~Kiritsis, L.~Mazzanti and F.~Nitti,
  Phys.\ Rev.\ Lett.\  {\bf 101}, 181601 (2008)
  [arXiv:0804.0899 [hep-th]].

\bibitem{Gubser:2008ny}
  S.~S.~Gubser and A.~Nellore,
  Phys.\ Rev.\ D {\bf 78}, 086007 (2008)
  [arXiv:0804.0434 [hep-th]].
  
\bibitem{Noronha:2009ud} 
  J.~Noronha,
  Phys.\ Rev.\ D {\bf 81}, 045011 (2010)
  [arXiv:0910.1261 [hep-th]].

\bibitem{Ali-Akbari:2014vpa} 
  M.~Ali-Akbari, D.~Giataganas and Z.~Rezaei,
  arXiv:1406.1994 [hep-th].
  
\end{thebibliography}
\end{document}